\documentclass[%
 reprint,
superscriptaddress,
showpacs,preprintnumbers,
 amsmath,amssymb,
 prb,
]{revtex4-1}

\usepackage{graphicx}
\usepackage{dcolumn}
\usepackage{bm}
\usepackage{xcolor}

\begin{document}

\newcommand{\Rb}{RbFe$_2$As$_2$}
\newcommand{\Cs}{CsFe$_2$As$_2$}
\newcommand{\K}{KFe$_2$As$_2$}
\newcommand{\Tc}{$T_{\rm c}$}
\newcommand{\Tn}{$T_{\rm N}$}
\newcommand{\Pc}{$P_{\rm c}$}
\newcommand{\PH}{$P_{\rm H}$}
\newcommand{\dhdt}{$\left(-\partial{\rm H_{c2}}/\partial{\rm T} \right)_{T_{c\rm}}$}
\newcommand{\ie}{{\it i.e.}}
\newcommand{\eg}{{\it e.g.}}
\newcommand{\etal}{{\it et al.}}  
\newcommand{\mucm}{$\mathrm{\mu\Omega\, cm}$}
\newcommand{\Tmin}{$T_{\rm min}$}
\newcommand{\Tstar}{$T^\star$}
\newcommand{\Hvs}{$H_{\rm vs}$}
\newcommand{\Hc}{$H_{\rm c2}$}
\newcommand{\Hcstar}{$H_{\rm c2}^\star$}
\newcommand{\Hstar}{$H^\star$}
\newcommand{\Nqp}{$N_{\rm qp}$}
\newcommand{\Nsc}{$N_{\rm sc}$}
\newcommand{\NbSi}{Nb$_{0.15}$Si$_{0.85}$}
\newcommand{\RH}{$R_{\rm H}$}
\newcommand{\xc}{$x_{\rm c}$}
\newcommand{\LB}{$\ell_{\rm B}$}


\title{Universal V-shaped temperature-pressure phase diagram in the iron-based superconductors KFe$_2$As$_2$, RbFe$_2$As$_2$, and CsFe$_2$As$_2$}

\author{F.~F.~Tafti}
\email{Fazel.Fallah.Tafti@USherbrooke.ca}
\affiliation{D\'epartement de physique \& RQMP, Universit\'e de Sherbrooke, Sherbrooke, Qu\'{e}bec, Canada}

 \author{A. Ouellet}%
\affiliation{D\'epartement de physique \& RQMP, Universit\'e de Sherbrooke, Sherbrooke, Qu\'{e}bec, Canada}%

\author{A. Juneau-Fecteau}
\affiliation{D\'epartement de physique \& RQMP, Universit\'e de Sherbrooke, Sherbrooke, Qu\'{e}bec, Canada}

\author{S. Faucher}
\affiliation{D\'epartement de physique \& RQMP, Universit\'e de Sherbrooke, Sherbrooke, Qu\'{e}bec, Canada}

\author{M. Lapointe-Major}
\affiliation{D\'epartement de physique \& RQMP, Universit\'e de Sherbrooke, Sherbrooke, Qu\'{e}bec, Canada}

\author{N.~Doiron-Leyraud}
\affiliation{D\'epartement de physique \& RQMP, Universit\'e de Sherbrooke, Sherbrooke, Qu\'{e}bec, Canada}

\author{A. F. Wang}%
\affiliation{Hefei National Laboratory for Physical Sciences at Microscale and Department of Physics, University of Science and Technology of China, Hefei, China}%

\author{X.-G. Luo}%
\affiliation{Hefei National Laboratory for Physical Sciences at Microscale and Department of Physics, University of Science and Technology of China, Hefei, China}%

\author{X. H. Chen}%
\affiliation{Hefei National Laboratory for Physical Sciences at Microscale and Department of Physics, University of Science and Technology of China, Hefei, China}%

\author{Louis Taillefer}
\email{Louis.Taillefer@USherbrooke.ca}
\affiliation{D\'epartement de physique \& RQMP, Universit\'e de Sherbrooke, Sherbrooke, QC, Canada}
\affiliation{Canadian Institute for Advanced Research, Toronto, ON, Canada}

\date{\today}

\begin{abstract}
We report a sudden reversal in the pressure dependence of \Tc~in the iron-based superconductor \Rb, at a critical pressure \Pc~=~11~kbar. 
Combined with our prior results on \K~and \Cs,
we find a universal V-shaped phase diagram for \Tc~vs $P$ in these fully hole-doped 122 materials,
when measured relative to the critical point (\Pc, \Tc).
%
%
%
From measurements of the upper critical field \Hc$(T)$ under pressure in \K~and \Rb, 
we observe the same two-fold jump in $(1/T_{\rm c})$\dhdt~across  \Pc,
compelling evidence for a sudden change in the structure of the superconducting gap.
We argue that this change is due to a transition from one pairing state to another, with different symmetries on either side of \Pc.
We discuss a possible link between scattering and pairing, and a scenario where a $d$-wave state favoured
by high-$Q$ scattering at low pressure changes to a state with $s_\pm$ symmetry favoured by low-$Q$ scattering at high pressure.
%
%
%
\end{abstract}

\maketitle


\section{\label{Introduction}Introduction}

Pairing symmetry in the iron based superconductors is controlled by the interband and the intraband interactions which are tunable by external parameters such as doping or pressure. \cite{chubukov_pairing_2012, hirschfeld_gap_2011}
Recent theoretical works using a five orbital tight binding model show a near degeneracy between $d$ and $s_\pm$ pairing states in the 122 iron pnictides as a result of the multiorbital structure of the Cooper pairs and the near nesting conditions. \cite{graser_near-degeneracy_2009, fernandes_suppression_2013}
Hydrostatic pressure is a clean tuning parameter that can modify the orbital overlap, the exchange interactions, and the band structure of metals.
Given the near degeneracy between different pairing states, it is conceivable that the pairing symmetry of certain iron-based superconductors can be tuned by pressure.
These ideas gained experimental relevance with our recent discovery of a V-shaped $T$-$P$ phase diagram in \K~and \Cs,
where \Tc~decreases initially as a function of pressure, then at a critical pressure \Pc, it suddenly changes direction and increases. \cite{tafti_sudden_2013, tafti_sudden_2014}
Our results on \K~have been reproduced by other groups. \cite{terashima_two_2014, taufour_upper_2014, grinenko_superconducting_2014}
We interpreted the transition at \Pc~as a change of pairing state, possibly from $d$ to $s_\pm$, where the decreasing \Tc~of the former meets the growing \Tc~of the latter at \Pc, resulting in a V-shaped phase diagram.
%

In this article, we present our discovery of a very similar V-shaped phase diagram in a third material: \Rb.
We show that $\partial T_{\rm c} / \partial P$ on both sides of \Pc~ is the same in all three materials, which therefore share a universal V-shaped phase diagram.
We show that the slope of \Hc$(T)$ jumps by a factor 2 across \Pc, in a manner that also appears to be universal.
This is compelling evidence for a sudden change in the gap structure, \cite{taufour_upper_2014} which we attribute to a change of pairing symmetry.

%

%
The Fermi surface of \K~is known in  detail from quantum oscillations studies. \cite{terashima_fermi_2010, terashima_fermi_2013} 
It consists of three quasi-2D hole-like cylinders centered on the $\Gamma$~point, 
labelled $\alpha$ (small inner cylinder), $\xi$ (middle cylinder), and $\beta$ (large outer cylinder), 
as well as small cylindrical satellites at the corners of the Brillouin zone, labelled $\varepsilon$. 
%
There is also one small 3D hole pocket at the Z point of the Brillouin zone. \cite{diego_a._zocco_fermi_2014}
The pairing symmetry of \K~at ambient pressure is the subject of ongoing debate. 
Laser ARPES experiments suggest a $s_\pm$ state with eight line nodes on the $\xi$ band and no nodes on the $\alpha$ and $\beta$ bands. \cite{okazaki_octet-line_2012, watanabe_doping_2014}
%
%
By contrast, bulk measurements of thermal conductivity, \cite{reid_universal_2012, reid_d-wave_2012}, heat capacity \cite{abdel-hafiez_evidence_2013, grinenko_superconducting_Cv_2014} and penetration depth \cite{hashimoto_evidence_2010, kim_evolution_2014} point to a $d$-wave state, with four line nodes on each of the $\Gamma$-centered Fermi pockets ($\alpha$, $\xi$, and $\beta$). 
%

%
%
%


\section{\label{Experiments}Methods}

Single crystals of \Rb~were grown using a self-flux method similar to \K~and \Cs. \cite{wang_calorimetric_2013, zhang_heat_2014}
Two samples of \Rb, labelled Sample A and Sample B, were pressurized in a clamp cell and measured up to 22~kbar.
The pressure was measured by monitoring the superconducting transition of a lead gauge placed besides the two samples in the clamp cell with a precision of $\pm~0.1$ kbar.
A 1/1 mixture of pentane and 3-­methyl-­1-butanol was used as the pressure medium.
Measurements of resistivity and Hall effect were performed on both samples in an adiabatic demagnetization refrigerator using the standard six-contact configuration.
Hall voltage was measured at plus and minus 10~T from $T=20$ to 1~K and antisymmetrized to calculate the Hall coefficient \RH. 
Excellent reproducibility was observed between the two samples.
To avoid redundancy, we present the results from Sample A, unless otherwise mentioned.


\section{\label{Results}Results and Discussion}

We have measured the pressure dependence of four quantities: 
the critical temperature \Tc, 
the Hall coefficient \RH,
the upper critical field \Hc$(T)$, 
and the resistivity $\rho$.  
We present each measurement in turn, and discuss the implications.

\subsection{\label{P_Tc}Pressure dependence of \Tc}

\begin{figure}
\includegraphics[width=3.5in]{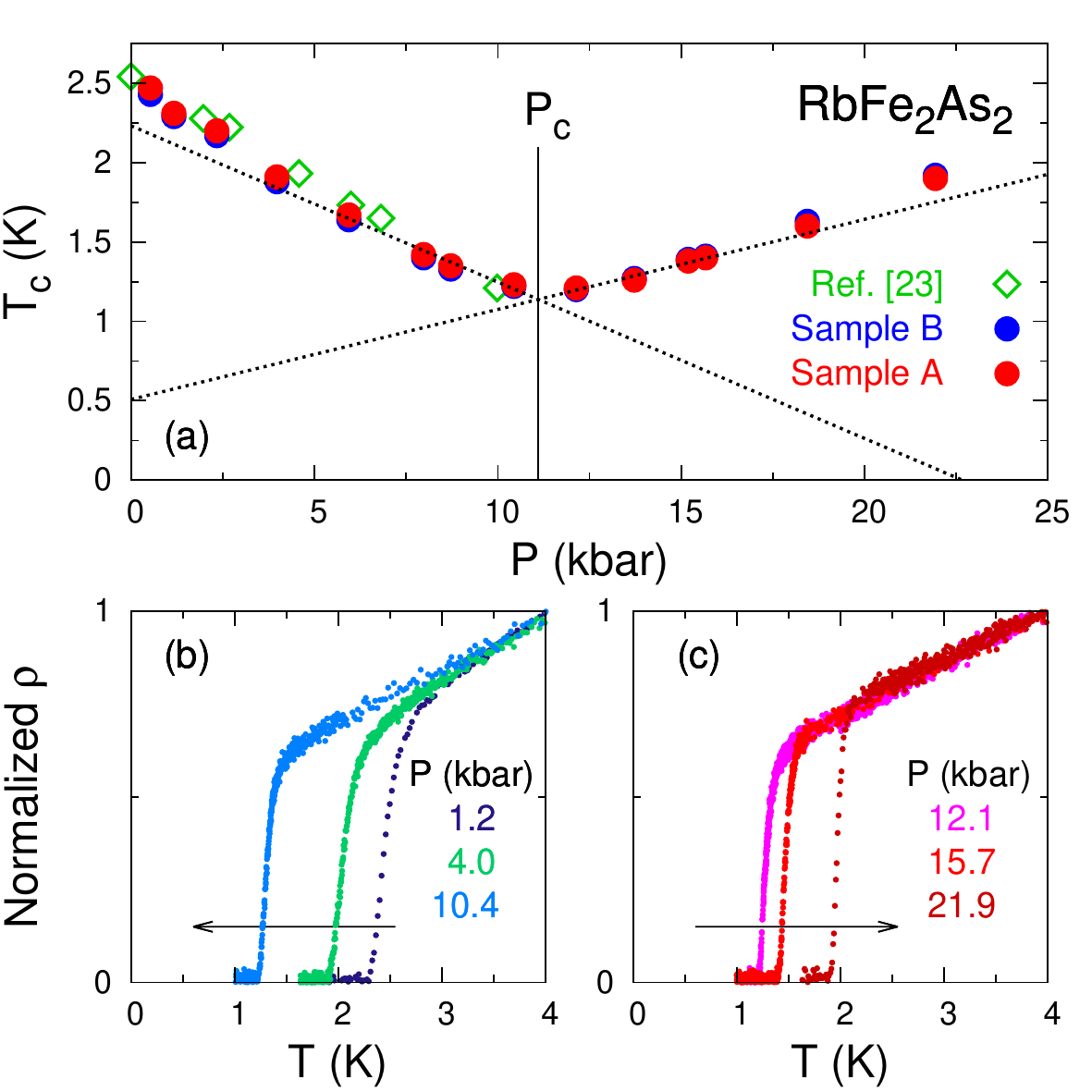}
\caption{\label{resistivity} 
(a) Pressure dependence of \Tc~in \Rb. 
The red and  blue circles represent data from Samples A and  B, respectively. 
\Tc~is defined as the temperature where $\rho=0$.
The critical pressure \Pc~marks the reversal in the \Tc~behaviour from decreasing to increasing. 
Dotted lines are linear fits to the data in the range \Pc~$\pm~6$ kbar. 
The critical pressure \Pc~$= 11 \pm 1$~kbar is defined as the intersection of the two dotted lines. 
The green diamonds are data from Ref.~[\onlinecite{shermadini_superfluid_2012}], where \Tc~was measured by AC susceptibility and $\mu$SR.
(b) Three representative $\rho(T)$ curves in the low pressure phase ($P < P_{\rm c}$), from Sample A, normalized to unity at $T=4$~K. 
The arrow shows that \Tc~decreases with increasing pressure. 
(c) Same as in (b) for the high pressure phase ($P > P_{\rm c}$).
The arrow shows that \Tc~now {\it increases} with increasing pressure.
}
\end{figure}

Figure~\ref{resistivity}(a) shows our discovery of a sudden reversal in the pressure dependence of \Tc~in \Rb~at a critical pressure \Pc~=~$11\pm 1$~kbar.
On the same figure, the green diamonds show \Tc~values reported previously for \Rb~from AC susceptiblity and $\mu$SR measurements up to 10 kbar by Shermadini \etal \cite{shermadini_superfluid_2012}
These data are in excellent agreement with ours, but they stop at 10~kbar, just before \Pc. 
By extrapolation, Shermadini \etal~concluded that \Tc~would be fully suppressed at 19~kbar which is the natural expectation from a superconductor with one dominant pairing state.
%
%

We determined \Tc~from resistivity measurements using a $\rho = 0$ criterion.
Figure~\ref{resistivity}(b) shows three representative resistivity curves at $P<P_c$, where \Tc~decreases by increasing pressure. 
Figure~\ref{resistivity}(c) shows three curves at $P>P_c$, where \Tc~reverses direction to increase by increasing pressure.
\Tc~varies linearly near \Pc~from either side, resulting in a $V$-shaped phase diagram in \Rb, similar to \K~and \Cs. \cite{tafti_sudden_2013, tafti_sudden_2014}

\begin{figure}
\includegraphics[width=3.5in]{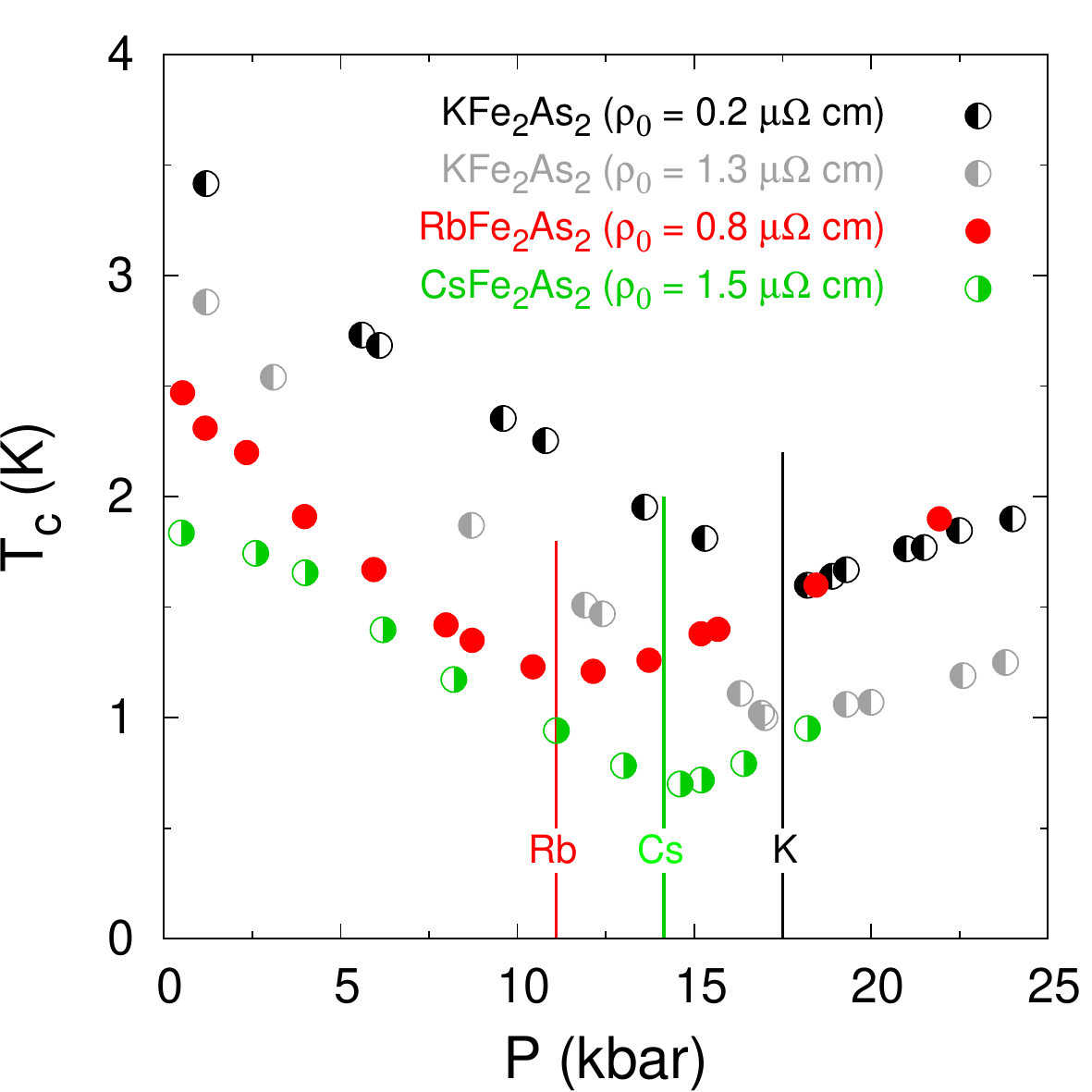}
\caption{\label{PhaseDiagram} 
Temperature-pressure phase diagram of \K~(black and grey), \Rb~(red), and \Cs~(green).
The residual resistivity  $\rho_0$ of each sample is indicated.
%
Comparing the black and  grey points shows that \Pc~is unaffected by disorder.
%
%
The three compounds show a \Tc~reversal at comparable critical pressures (color-coded vertical lines):
\Pc~=~17.5, 11, and 14 kbar, in \K, \Rb~and \Cs, respectively.
%
%
Data for \K~and \Cs~are reproduced from Refs.~\onlinecite{tafti_sudden_2013} and~\onlinecite{tafti_sudden_2014}, respectively.
Error bars are no larger than the size of the points.
}
\end{figure}

In Fig.~\ref{PhaseDiagram}, we compare the phase diagrams of \K, \Rb~and \Cs, with \Pc~= 17.5, 11, and 14 kbar, respectively.
%
%
The black and grey symbols correspond to two samples of \K~with different disorder levels.
%
%
Although the residual resistivity of the less pure sample ($\rho_0 = 1.3~\mu \Omega$ cm) is six times larger than that of the pure sample ($\rho_0 = 0.2~\mu \Omega$ cm), they have identical \Pc.
Hence, disorder does not affect the critical pressure.

\begin{table*}
\caption{\label{lattice} 
Lattice parameters ($a$ and $c$), unit cell volume ($V=ca^2$),  Sommerfeld coefficient in the specific heat ($\gamma$),  critical temperature \Tc~at ambient pressure, \Tc~at~\Pc, and critical pressure \Pc, 
for \K, \Rb~and \Cs. \cite{sasmal_superconducting_2008, aftabuzzaman_new_2010, hardy_evidence_2013, wang_calorimetric_2013, zhang_heat_2014}
As the atomic size of the alkali ion increases from K to Rb to Cs, lattice parameters, unit cell volume, and the Sommerfeld coefficient systematically increase, while \Tc~systematically decreases.
But the critical pressure \Pc~does not follow these systematics:
it decreases from K to Rb, then increases from Rb to Cs.
}
\begin{ruledtabular}
\begin{tabular}{lccccccc}
Material & $a$~($\rm{\AA}$)  & $c$~($\rm{\AA}$) &  $V$~($\rm{\AA}^3$) & $\gamma$~(mJ/K$^2$mol)  & \Tc~($P=0$) (K) & \Tc~($P=$ \Pc) (K)  & \Pc~(kbar)  \\
\colrule 
\K   & 3.84 & 13.84 & 204  & 99 & 3.6 & 1.6 & 17.5 \\
\Rb  & 3.86 & 14.45 & 215  & 128 & 2.5 & 1.2 & 11 \\
\Cs  & 3.89 & 15.07 & 228 & 184 & 1.8 & 0.7  & 14

\end{tabular}
\end{ruledtabular}
\end{table*}

\begin{figure}[t]
\includegraphics[width=3.5in]{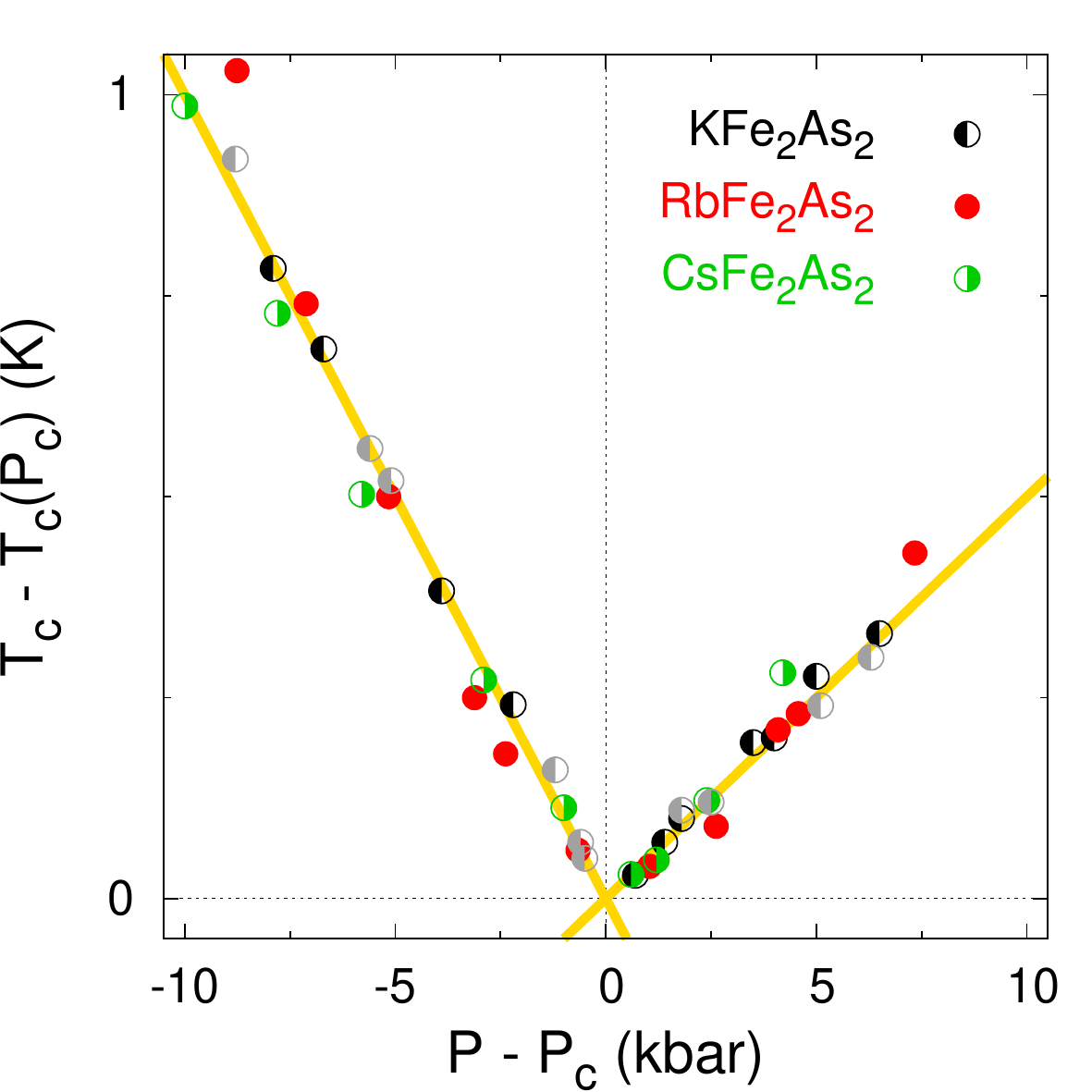}
\caption{\label{PhaseDiagram_Shifted}
Universal V-shaped phase diagram of AFe$_2$As$_2$, where A = K (black and grey), Rb (red) and Cs (green).
Data points from Fig.~\ref{PhaseDiagram}
are re-plotted with pressure values shifted by the respective critical pressures (\Pc~=~17.5, 11, and 14 kbar, for A = K, Rb, and Cs)
and \Tc~values shifted by the respective values of \Tc~at \Pc~(1.6, 1.2, and 0.7~K, for A = K, Rb, and Cs).
%
%
%
Error bars are no larger than the size of the points.
The yellow lines are linear fits on either side of \Pc.
}
\end{figure}

Figure~\ref{PhaseDiagram} also shows that the phase diagram of the less pure \K~sample (gray) is rigidly shifted down relative to the pure sample (black).
Therefore, the low-pressure and the high-pressure superconducting phases have comparable (and large) sensitivity to disorder.
The high sensitivity to disorder on both sides of \Pc~is incompatible with the standard $s_{++}$ state,
known to be resilient against disorder. \cite{anderson_theory_1959, prozorov_effect_2014}
In \K~at ambient pressure, the steep drop of \Tc~as a function of controlled disorder was shown to be quantitatively consistent with the high sensitivity of a $d$-wave state to disorder. \cite{wang_anomalous_2014, reid_universal_2012, kirshenbaum_universal_2012}
Since the usual $s_\pm$ state, with a sign change between pockets centered at different points in the Brillouin zone, is known to be much less sensitive to disorder than the $d$-wave state, \cite{kirshenbaum_universal_2012}
it may be that the high pressure phase is the $s_\pm^h$ state proposed  for \K, \cite{maiti_gap_2012}
where the gap changes sign between two of the $\Gamma$-centered pockets resulting in a higher sensitivity to disorder. 
%

\begin{figure}
\includegraphics[width=3.5in]{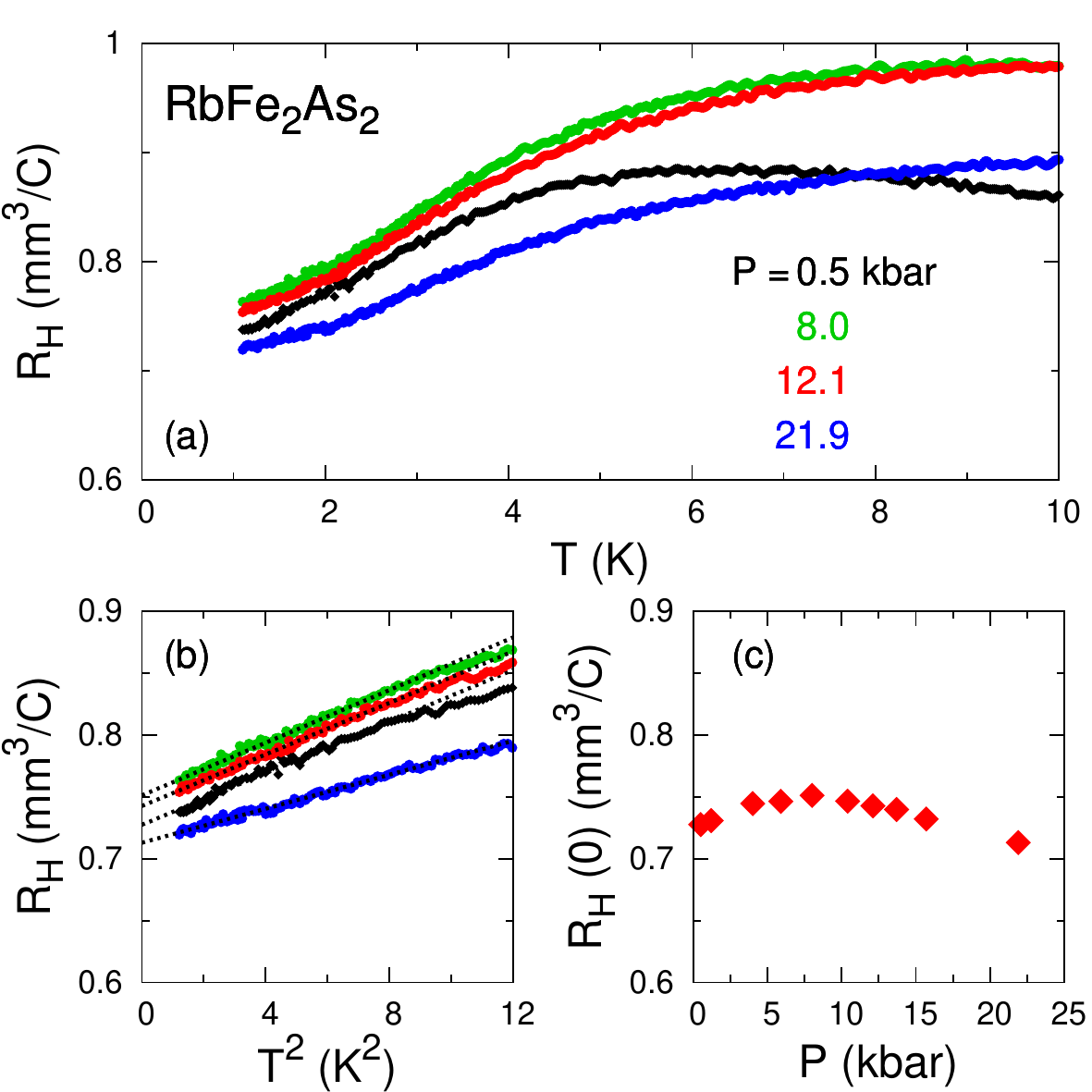}
\caption{\label{hall} 
(a) Temperature dependence of the Hall coefficient \RH$(T)$ in \Rb~at four pressures as indicated. 
%
(b) \RH$(T)$ plotted versus $T^2$ for the same pressures as in (a), at low temperature.
The dotted lines are linear fits from which the zero-temperature limit of the Hall coefficient, \RH$(0)$,  is obtained at each pressure. 
(c) \RH$(0)$ plotted as a function of pressure, showing a small but smooth variation.
Vertical and horizontal error bars are no larger than the size of the points.
Vertical error bars come from the uncertainty in the $T^2$ fits in (b), associated with changing the fitting interval.
}
\end{figure}

%
By measuring the lattice parameters of \K~under pressure, we showed in Ref.~[\onlinecite{tafti_sudden_2014}] that it takes about 30~kbar 
to tune the unit cell volume and the As-Fe-As bond angle of \Cs~to match those of \K.
Therefore, if the structural parameters directly controlled~\Pc, one would expect the critical pressure of \Cs~to be 30~kbar larger than that of~\K.
As shown in Fig.~\ref{PhaseDiagram}, the~\Pc~of \Cs~is {\it less} than the~\Pc~of \K.
Table~\ref{lattice} lists the structural parameters of the three compounds at ambient pressure.
Rb atoms are intermediate in size between K and Cs, hence the lattice parameters of \Rb~are in between those of \K~and \Cs.
Fig.~\ref{PhaseDiagram} shows that the \Pc~of \Rb~does not fall between those of the other two compounds.
There is indeed no straightforward connection between lattice parameters, bond angles, or ionic sizes and the actual values of \Pc~in these three compounds.
Interestingly, near optimal doping, in both 122 and 1111 families, there \emph{is} indeed a straightforward connection between the structural parameters and \Tc. \cite{rotter_superconductivity_2008, miyoshi_superconductivity_2013} 
Near optimal doping, $d$ wave is known to be a subdominant superconducting state while $s$ wave is the dominant one. \cite{kretzschmar_raman-scattering_2013} 
The Loss of this straightforward connection between superconductivity and structural parameters in the over-doped regime could in fact result from a dominant $d$ wave state.
In Fig.~\ref{PhaseDiagram_Shifted}, we make a direct comparison of the three V-shaped phase diagrams, by shifting each curve horizontally 
by \Pc~and vertically by \Tc($P$=\Pc), so that the tip of the V coincides for the three materials.
Remarkably, we find the same, universal V-shaped phase diagram in the range \Pc~$\pm$ 10~kbar.
In other words, $(\partial T_{\rm c} / \partial P)_{P < P_{\rm c}}$ and $(\partial T_{\rm c} / \partial P)_{P>P_{\rm c}}$ are identical in the three materials.
%
%
Such a universal phase diagram, independent of the alkali atom (A) in AFe$_2$As$_2$ and of the different structural parameters, poses a clear challenge to our understanding of multiband superconductivity. 


\begin{figure}
\includegraphics[width=3.5in]{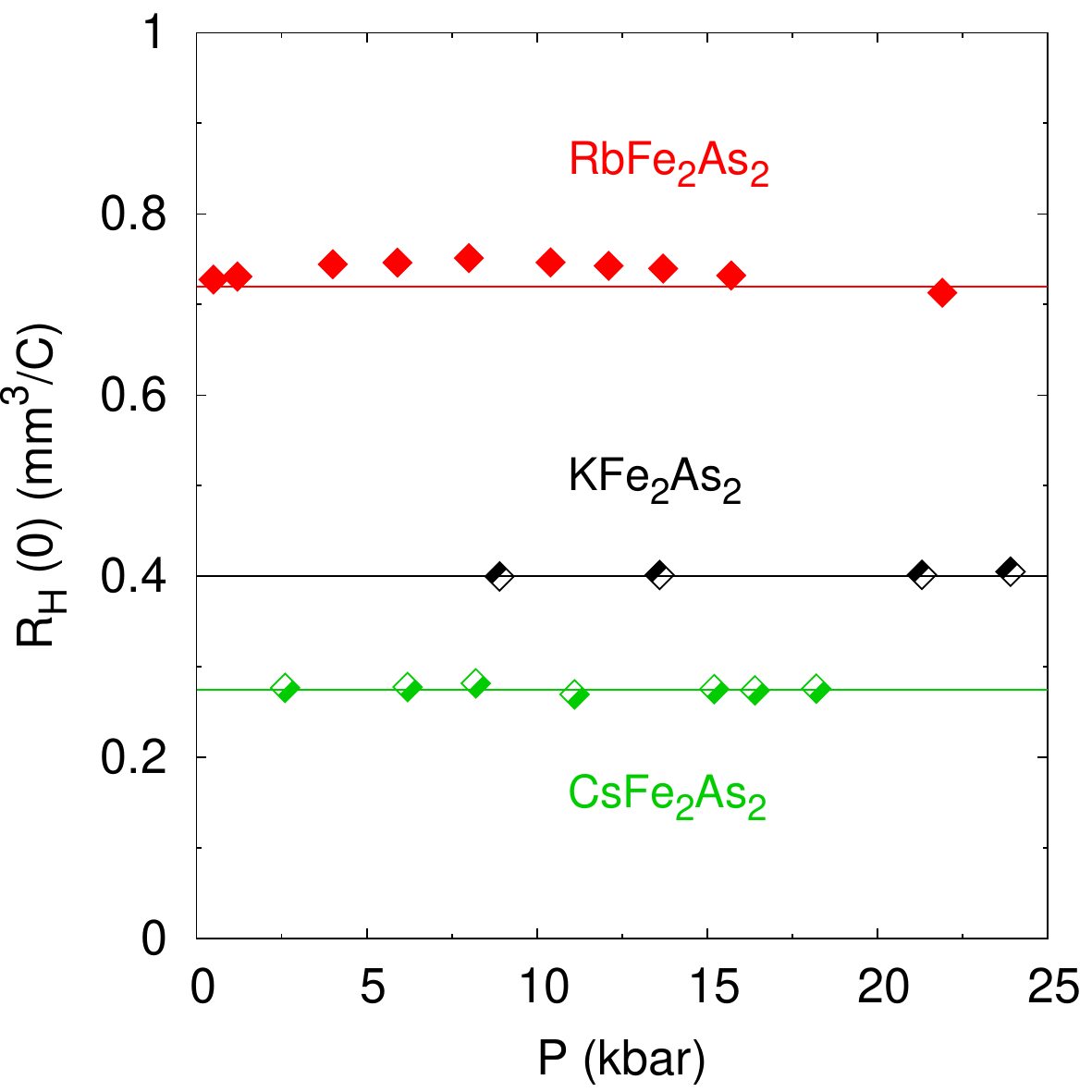}
\caption{\label{RH_0vsP} 
The zero-temperature limit of the Hall coefficient, \RH(0), plotted for the three materials 
AFe$_2$As$_2$, where A = K (black), Rb (red), and Cs (green).
Vertical and horizontal error bars are no larger than the size of the points.
Within error bars, \RH(0) is constant for \K~and \Cs, and nearly so for \Rb.
Horizontal lines are a guide to the eye.
Data for \K~and \Cs~are reproduced from Refs.~\onlinecite{tafti_sudden_2013} and~\onlinecite{tafti_sudden_2014}, respectively.
}
\end{figure}

\subsection{\label{P_RH}Pressure dependence of \RH}

%
In our previous work on \K~and \Cs, we showed that the V-shaped pressure dependence of \Tc~is not accompanied by any abrupt changes in the normal-state properties.
In particular, the zero-temperature limit of the Hall coefficient, \RH$(0)$, was found constant in both materials as a function of pressure across \Pc, ruling out 
a sudden change in the Fermi surface, i.e. a Lifshitz transition. \cite{tafti_sudden_2013, tafti_sudden_2014}
We concluded that the phase transition at \Pc~is not triggered by some change in the Fermi surface, and is instead associated with a change in the superconducting state itself.
This was confirmed by quantum oscillation measurements on \K~under pressure, which found a smooth evolution of 
the oscillation frequencies and cyclotron masses across \Pc. \cite{terashima_two_2014}
%

\begin{figure}[t]
\includegraphics[width=3.5in]{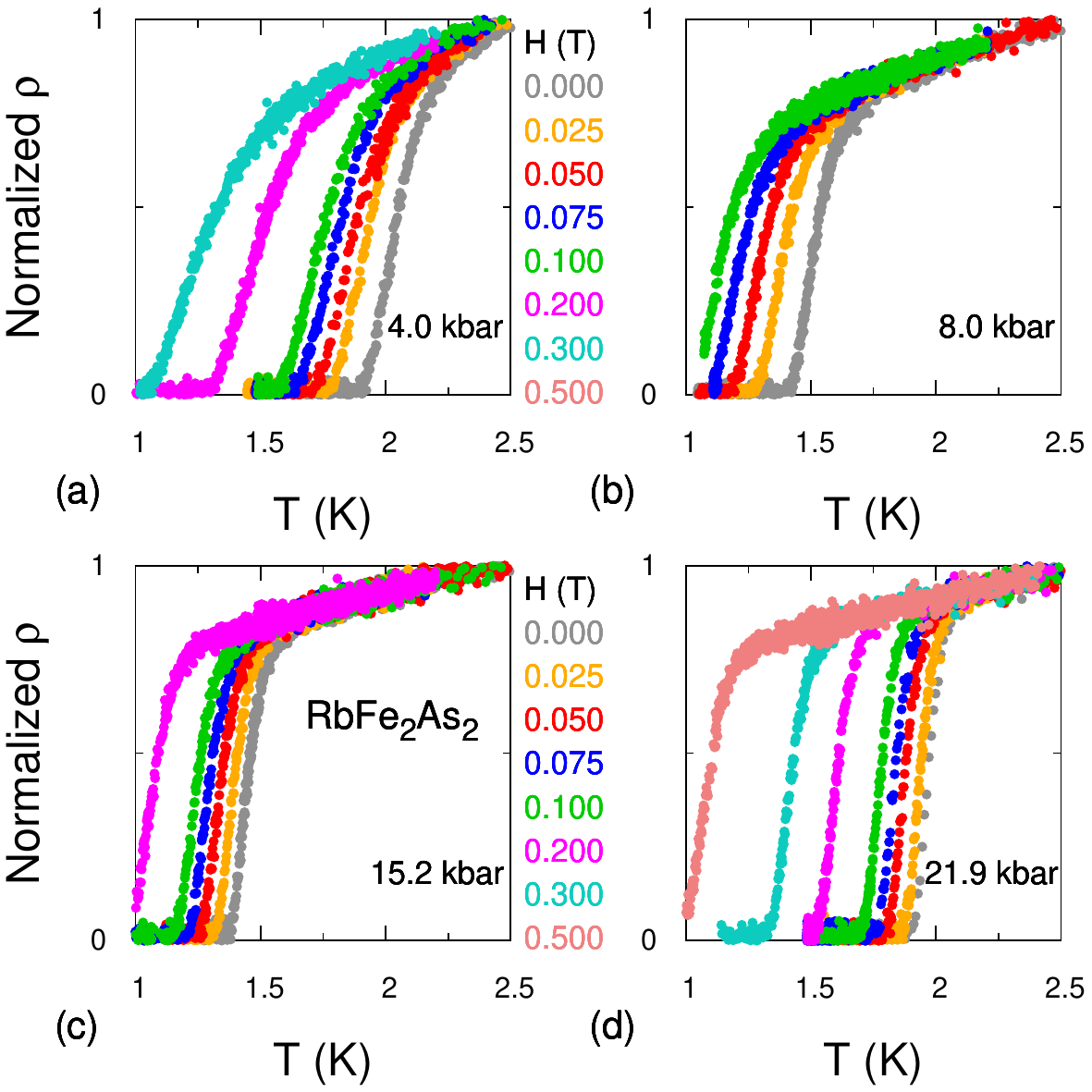}
\caption{\label{ResistiveHc2Rb}
Resistivity $\rho(T)$ of \Rb~normalized to unity at $T=2.5$~K, at $P = 4.0$, 8.0, 15.2, and 21.9~kbar.
The color of each curve corresponds to a field $H$ as indicated.
At each field, \Tc~is defined from the resistive transition using the criterion  $\rho = 0$.
This yields the curves of \Hc~vs $T$ in Fig.~\ref{Hc2}(a).
Note that the width of the transition is larger in the low-pressure phase (e.g. in (a)) compared to the high-pressure phase (e.g. in (d)),
because $\partial T_{\rm c} / \partial P$ is almost two times  higher in the low-pressure phase (Fig.~\ref{PhaseDiagram_Shifted}).
Therefore, a small pressure gradient across the sample in the pressure cell generates a wider transition in the low-pressure phase.
}
\end{figure}

Figure~\ref{hall}(a) shows the temperature dependence of \RH~in \Rb, at various pressures.
The low-temperature data goes as $T^2$, allowing us to extract \RH$(0)$ from a linear fit of \RH$(T)$ vs $T^2$, as shown in Fig.~\ref{hall}(b).
The \RH$(0)$ values are plotted as a function of pressure in Fig.~\ref{hall}(c).
We find a small but smooth variation in \RH$(0)$, with no anomaly at \Pc~=~11~kbar,
very different from the step-like structure expected of a typical Lifshitz transition. \cite{liu_evidence_2010}

In Figure~\ref{RH_0vsP}, we summarize the results of our Hall measurements in the three materials.
In \K~and \Cs, \RH$(0)$ is completely flat with pressure.
In \Rb, there is a broad structure in \RH$(0)$ and a straight line does not go through all data points.
We attribute this broad feature to a smooth evolution of Fermi surface parameters under pressure.
Taken together, these Hall data make a compelling case that the universal V-shaped \Tc~vs $P$ curve of 
Fig.~\ref{PhaseDiagram_Shifted} is not shaped by a sudden change in the Fermi surface at \Pc.

\begin{figure}
\includegraphics[width=3.5in]{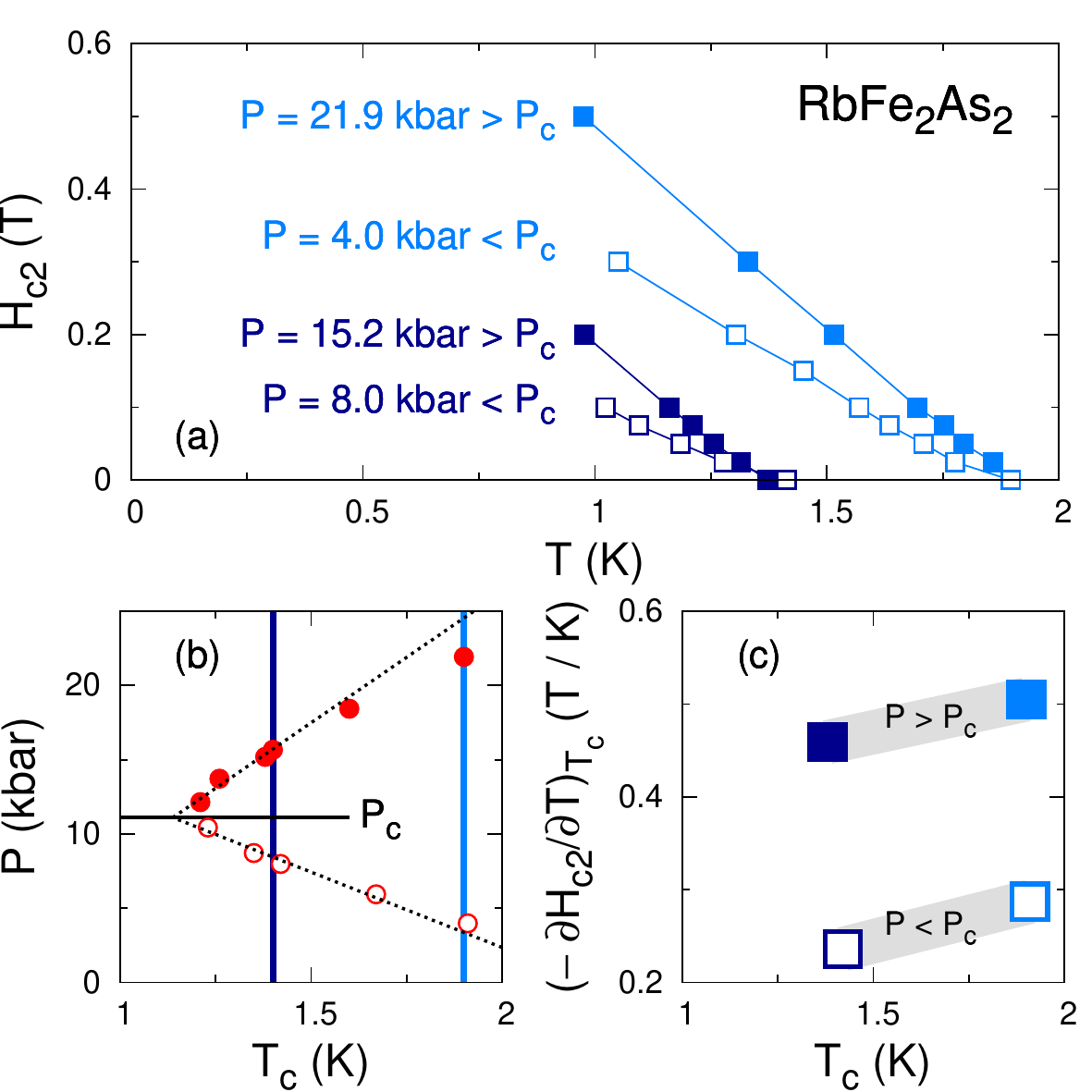}
\caption{\label{Hc2} 
The full symbols in this figure represent the high-pressure phase ($P>$ \Pc) and the empty symbols represent the low-pressure phase ($P<$ \Pc).
(a) \Hc$(T)$~in \Rb~at four pressures.
The light blue pair of curves starts from \Tc~$\simeq$~1.9~K and the dark blue pair starts from \Tc~$\simeq$~1.4~K, at $H = 0$. 
At a given \Tc, the low pressure curve has a lower slope compared to the high pressure one.
(b) The V-shape phase diagram of \Rb, rotated by 90 degrees, with lines that cut through the ``V" at pairs of pressures corresponding to the light blue and the dark blue curves in panel (a). 
%
(c) \dhdt~is extracted from linear fits to the curves in panel (a) near \Tc~and plotted against \Tc.
The light blue pair of points is vertically aligned at \Tc~$\simeq$~1.9 K and the dark blue pair at \Tc~$\simeq$~1.4 K corresponding to the vertical lines in panel (b).
Error bars are no larger than the size of the points.
We observe that consistently, \dhdt is larger in the high pressure phase.
}
\end{figure}

\subsection{\label{P_Hc2}Pressure dependence of \Hc}

%
Measurements of the upper critical field \Hc~in \K~under pressure can be used to provide evidence for a change of gap structure across \Pc.
Recently, Taufour \etal~found a change of regime in the dependence of the quantity $(1/T_{\rm c})$\dhdt~ as a function of the $A$ coefficient of resistivity from $\rho(T)=\rho_0+A T^2$.  \cite{taufour_upper_2014}
They linked this change of regime to a change in the $k_z$ corrugations of the superconducting gap at the critical pressure \Pc.
Inspired by their work, we explored the pressure dependence of \Hc~in both \K~and \Rb.
At any given pressure, we used measurements of $\rho(T)$ at different fields to determine \Hc($T$).
%
%
Figure~\ref{ResistiveHc2Rb} shows normalized $\rho(T)$ curves in \Rb~at different fields, for four representative pressures, below and above \Pc~=~11~kbar.
We define \Tc~at each field from the resistive transition using the $\rho = 0$ criterion, then plot those \Tc~values vs $H$ at each pressure, to arrive at the $H$-$T$ curves shown in Fig.~\ref{Hc2}(a).

\begin{figure}
\includegraphics[width=3.5in]{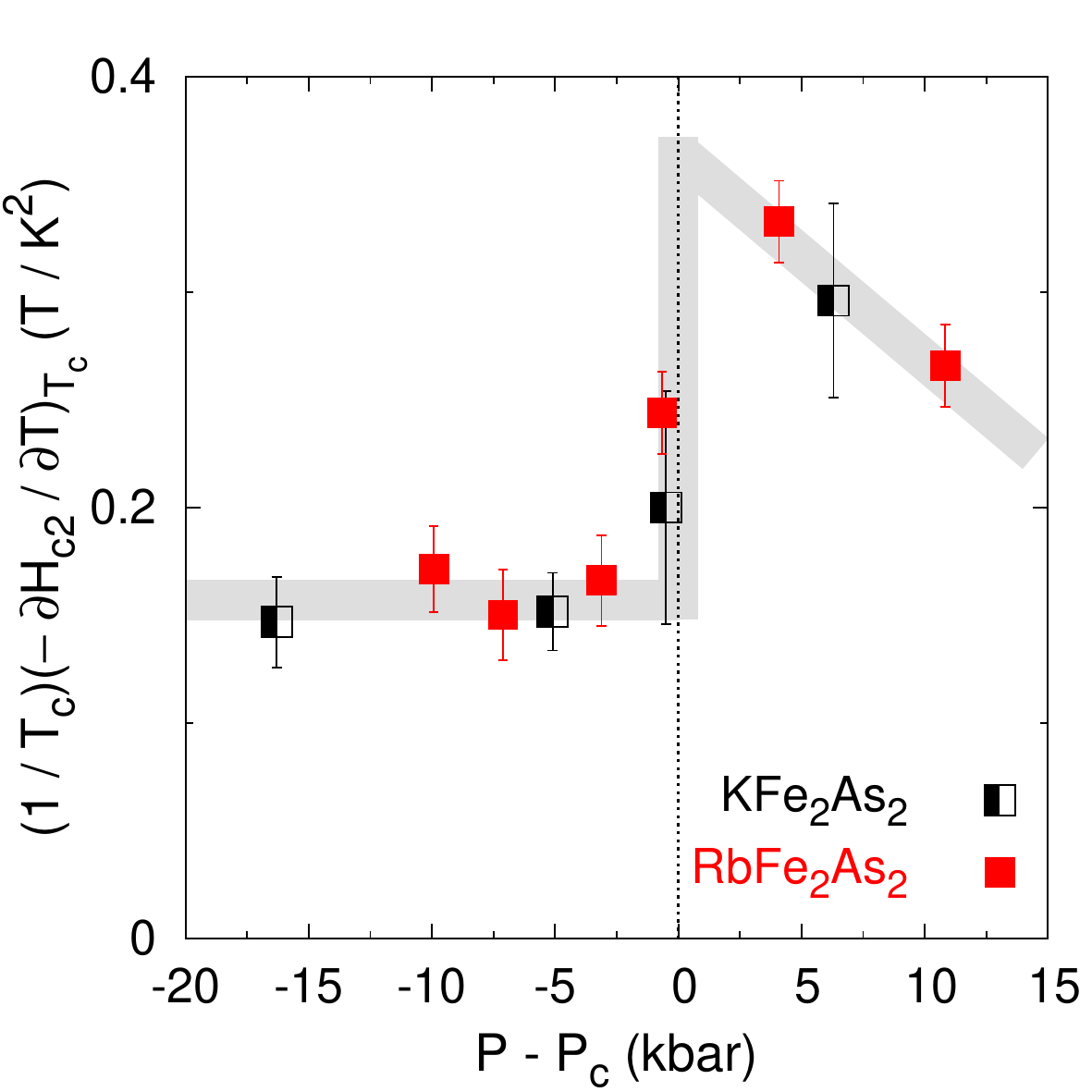}
\caption{\label{UniversalStep} 
$(1/T_{\rm c})$\dhdt~as a function of $P-$\Pc~in \Rb~(red squares) and \K~(black squares). 
The two materials are seen to display the same behaviour, with the same magnitude on both sides of \Pc.
The thick gray line is a guide to the eye.
A two-fold jump is observed at $P=$~\Pc, revealing a clear difference between the low-pressure phase ($P-P_{\rm c}<0$) and the high-pressure phase ($P-P_{\rm c}>0$).  
}
\end{figure}

%
The four \Hc$(T)$ curves in Fig.~\ref{Hc2}(a) should be viewed in two pairs with comparable \Tc~values on either side of \Pc, as illustrated in Fig.~\ref{Hc2}(b). 
%
%
%
%
%
%
%
Figure~\ref{Hc2}(c) summarizes the results by plotting \dhdt~versus \Tc~for the two pairs of points at \Tc~=~1.4~K and 1.9~K.
At each value of \Tc, the full symbol lies above the empty one, showing that the high-pressure phase is more robust against the magnetic field than the low-pressure phase. 
To obtain \dhdt~in \Rb, we made linear fits to the $H$-$T$ curves in Fig.~\ref{Hc2}(a) near \Tc.
%

\begin{figure}[t]
\includegraphics[width=3.5in]{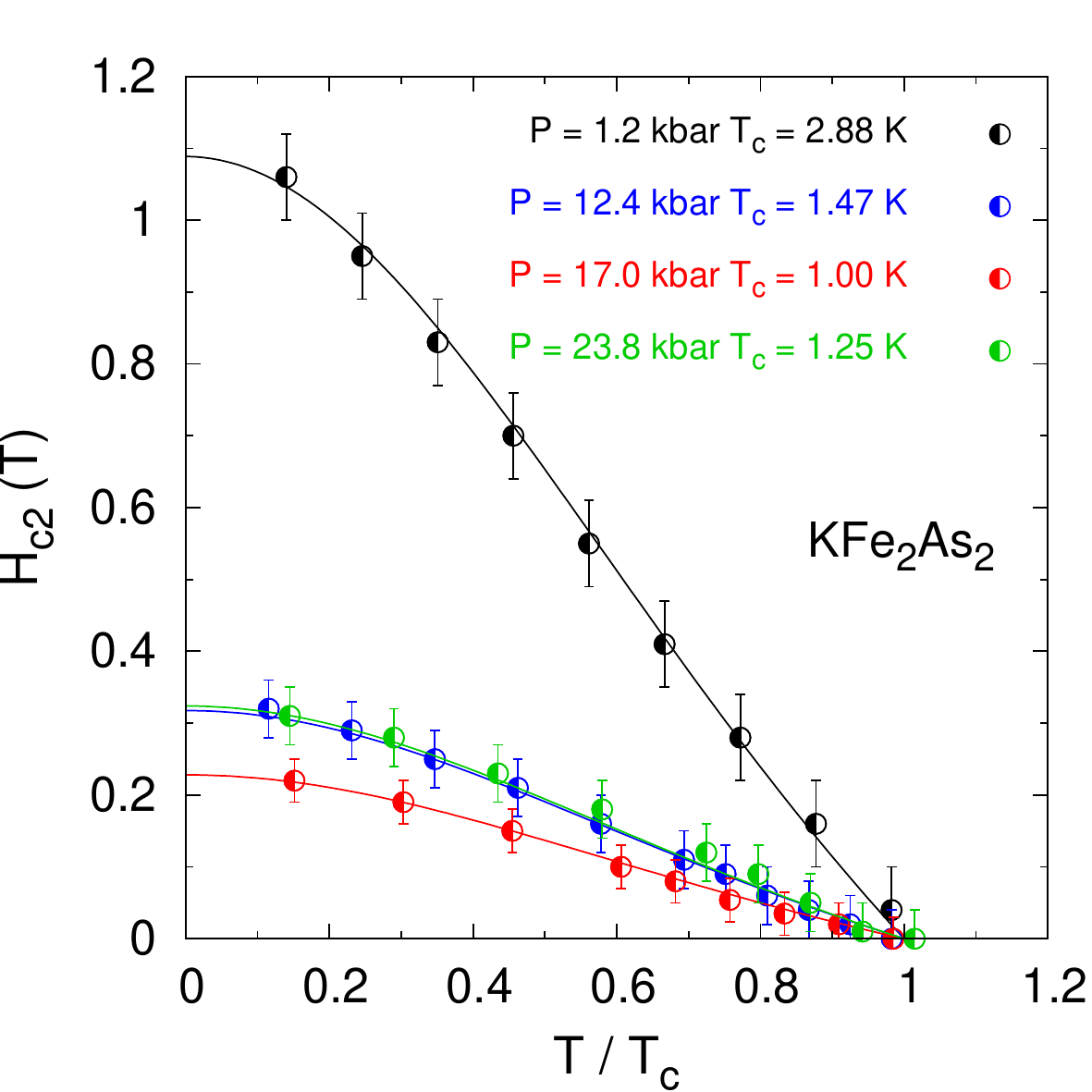}
\caption{\label{KF00_Hc2vst}
\Hc($T$) in \K~plotted as a function of $T/$\Tc.
At each pressure, \dhdt~is obtained by fitting a line to the \Hc$(T)$ curve near \Tc.
The resulting $(1/T_{\rm c})$\dhdt~values are the black data points in Fig.~\ref{UniversalStep}.
Solid lines are a guide to the eye.
}
\end{figure}

Figure~\ref{UniversalStep} shows $(1/T_{\rm c})$\dhdt~vs ($P-P_{\rm c}$) for both \Rb~and \K.
%
%
In total, we measured \Hc($T$)~at six different pressures, corresponding to the six red squares in Fig.~\ref{UniversalStep}.
The four black data points in Fig.~\ref{UniversalStep} come from similar measurements on \K, from the \Hc($T$)~curves shown in Fig.~\ref{KF00_Hc2vst}. 
The magnitude of $(1/T_{\rm c})$\dhdt~is the same in the two materials, on both sides of \Pc, revealing a second universal property of the pressure-induced transition:
%
%
a two-fold jump in $(1/T_{\rm c})$\dhdt~at $P=$~\Pc~(Fig.~\ref{UniversalStep}).
%
%
%


%
The Helfand-Werthamer theory, modified for superconductors with anisotropic gap structures, \cite{kogan_orbital_2012} relates the quantity $(1/T_c)$\dhdt~to the gap dispersion $\Omega$ and the anisotropic Fermi velocity $v_0$ through the relation:
\begin{equation}
\label{kogan}
\frac{1}{T_{\rm c}} \left( \frac{- \partial H_{\rm c2}}{\partial T} \right)_{T_{\rm c}} = \frac{16 \pi \phi_0 k_{\rm B}^2}{7 \zeta (3) \hbar^2} \frac{1}{\langle \Omega ^2 \mu_{\rm c} \rangle} \frac{1}{v_0 ^2}
\end{equation}
where $\phi_0$, $\hbar$, and $k_{\rm B}$ are the magnetic flux quantum, the Planck constant, and the Boltzmann constant. 
$\mu_{\rm c}$ and $v_0$ relate to the Fermi surface parameters according to:
\begin{equation}
\label{muc}
\mu_{\rm c} = \frac{v_x^2 + v_y^2}{v_0 ^2} \qquad {\rm and} \qquad v_0^3 = \frac{2 E_{\rm F}^2}{\pi^2 \hbar^3 N(0)} 
\end{equation}
Where $E_{\rm F}$ and $N(0)$ are the Fermi energy and the density of states at the Fermi level.
Fermi velocity is not a constant at anisotropic Fermi surfaces, hence, a characteristic constant velocity $v_0$ is defined in Eq.~\ref{muc} applicable to anisotropic Fermi surfaces. \cite{kogan_orbital_2012}
For the isotropic case, one recovers $v_0 = v_{\rm F}$.
The function $\Omega$ describes the momentum dependence of the superconducting gap: $\Delta = \Psi({\bf r},{\bf T}) \Omega({\bf k_{\rm F}})$.
The averages $\langle \cdots \rangle$ are taken over the Fermi surface.

Given that quantum oscillation measurements in \K~under pressure show a smooth evolution of $v_{\rm F}$ across \Pc, \cite{terashima_two_2014}
the two-fold jump in $(1/T_{\rm c})$\dhdt~observed at \Pc~can only be the result of a sudden change in the gap dispersion function $\Omega$ in Eq.~\ref{kogan},
caused by a sudden change in the structure of the superconducting gap across \Pc.
The most natural mechanism for such a change is a transition from one pairing state to another, with a change of symmetry, {\it e.g.} from $d$-wave to $s$-wave.

\subsection{\label{P_rho20}Pressure dependence of $\rho(T)$}

%

In a recent theoretical work, Fernandes and Millis showed how different spin-mediated pairing interactions 
in iron-based superconductors can favor different pairing symmetries. \cite{fernandes_suppression_2013}
In their model, based on the standard Fermi surface with hole-like pockets at the $\Gamma$ point and electron pockets at the $X$ points,
SDW-type magnetic fluctuations, with wavevector $(\pi,0)$,  favour  $s_{\pm}$ pairing, whereas
 N\'eel-type fluctuations, with wavevector $(\pi,\pi)$, favour $d$-wave pairing.
A gradual increase in the $(\pi,\pi)$ fluctuations rapidly suppresses the $s_{\pm}$ superconducting state and eventually causes a phase transition from the $s_{\pm}$ 
 state to a $d$-wave state, producing a V-shaped phase diagram of \Tc~vs tuning parameter (strength of $(\pi,\pi)$ correlations). \cite{fernandes_suppression_2013}

\begin{figure}[t]
\includegraphics[width=3.5in]{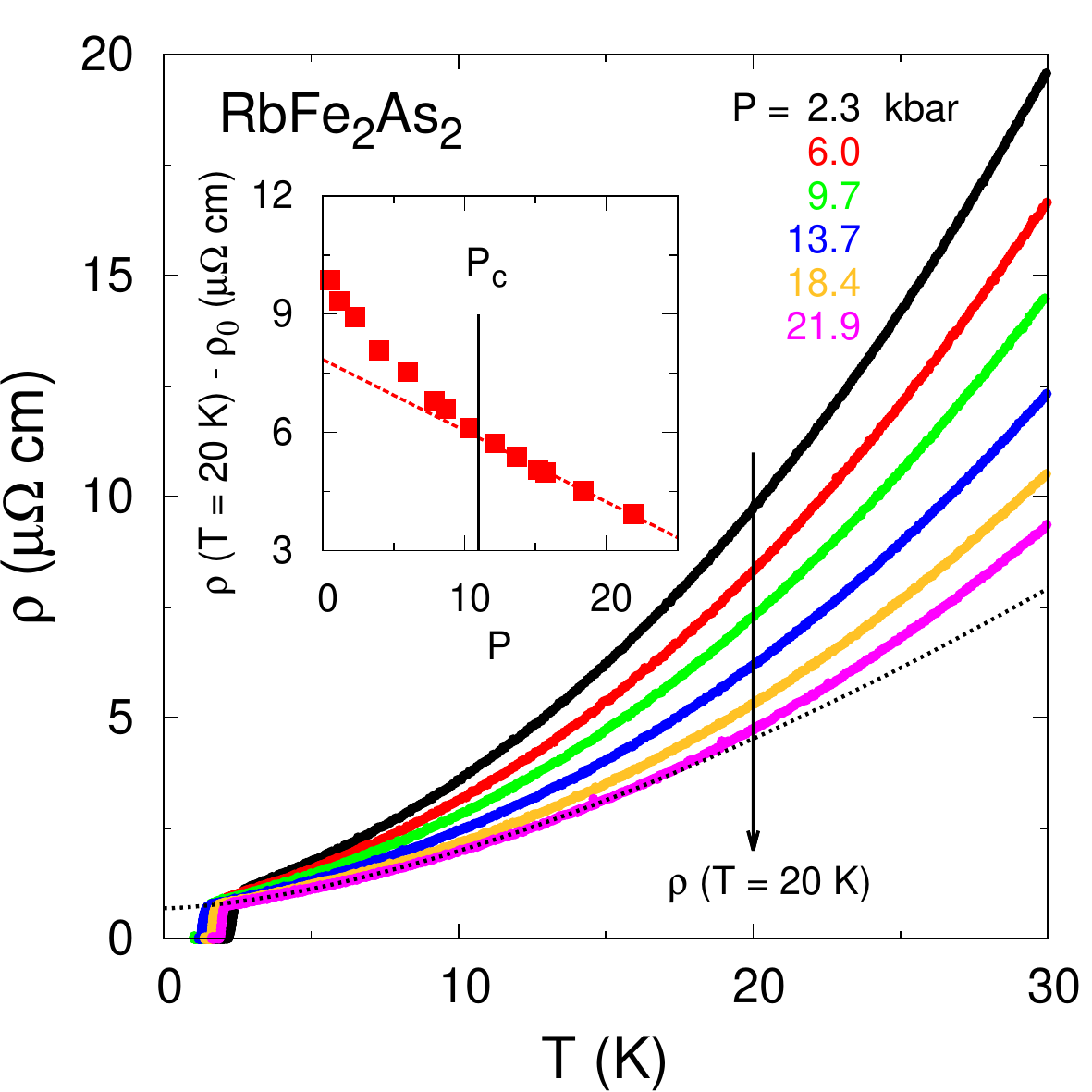}
\caption{\label{R20K}
Resistivity of sample A as a function of temperature up to 30 K, at six representative pressures, as indicated. 
The vertical arrow that cuts through the resistivity curves at $T=20$~K defines $\rho(T=20{\rm K})$. 
The black dotted line is a power law fit to the resistivity curve at $P=21.9$~kbar, in the interval from $T=3$~K to 15 K.
We define the residual resistivity $\rho_0$ as the zero temperature limiting value of this fit. 
The inset shows the pressure dependence of the inelastic resistivity, $\rho(T = 20~\rm{K}) - \rho_0$. 
The vertical arrow in the inset shows \Pc~=~11~kbar. 
The red dashed line is a linear fit to the inelastic resistivity as a function of pressure at $P>$~\Pc. 
}
\end{figure}

\begin{figure}[t]
\includegraphics[width=3.5in]{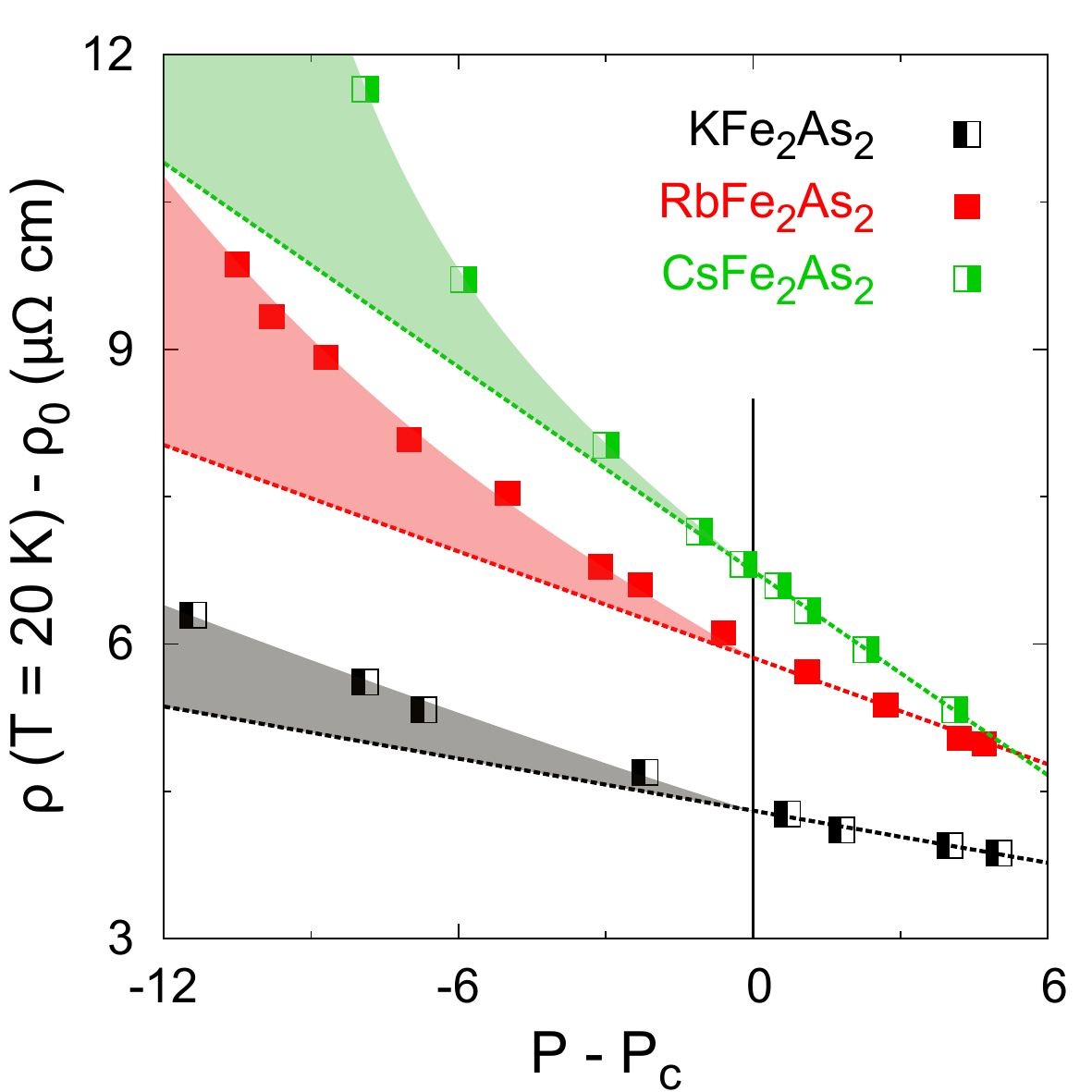}
\caption{\label{InelasticScattering_universal_Filled}
Inelastic resistivity defined as $\rho(T = 20~\rm{K}) - \rho_0$ and plotted versus $P - P_{\rm c}$ in \K~(black symbols,  \Pc~=~17.5~kbar), \Rb~(red symbols, \Pc~=~11~kbar), and \Cs~(green symbols, \Pc~=~14~kbar).
Dashed lines are linear fits to the data at $P - P_{\rm c} > 0$ showing that the inelastic resistivity scales linearly with pressure in the high pressure phase above \Pc.
As the pressure is lowered below \Pc, a new channel contributes to the inelastic resistivity of AFe$_2$As$_2$ shown as shaded areas in gray, red, and green for A = K, Rb, and Cs.
The vertical black line marks $P = P_{\rm c}$. 
}
\end{figure}

It is conceivable that two such competing interactions are at play in AFe$_2$As$_2$, with pressure tilting the balance in favor of one versus the other.
We explore such a scenario by looking at how the inelastic scattering evolves with pressure, measured via the inelastic
resistivity, defined as $\rho(T=20{\rm K}) - \rho_0$.
Figure~\ref{R20K} shows the temperature dependence of resistivity, at six representative pressures, in \Rb.
The residual resistivity $\rho_0$ comes from a power-law fit to the resistivity at any pressure, of the form $\rho=\rho_0+AT^n$.
The resulting $\rho_0$ does not change with pressure.
$\rho(T=20{\rm K})$ is the value of the resistivity at $T=20$~K, at any given pressure.
Through the same procedure, we extracted $\rho_0$ and $\rho(T=20{\rm K})$ as a function of pressure in \K~and \Cs, from published data. \cite{tafti_sudden_2013, tafti_sudden_2014}
%
%
In Fig.~\ref{InelasticScattering_universal_Filled}, the inelastic resistivity $\rho(T= \rm{20K})-\rho_0$ is plotted as a function of $P - $\Pc. 
In all three materials, at $P - P_{\rm c} > 0$, the inelastic resistivity varies linearly with pressure.
As $P$ drops below \Pc~in the low pressure phase ($P - P_{\rm c} < 0$),  the inelastic resistivity shows a clear rise over and above the linear regime.
Figure \ref{InelasticScattering_universal_Filled} therefore suggests a connection between the transition in the pressure dependence of \Tc~and 
the appearance of an additional inelastic scattering channel.
Note that our choice of $T=20$~K to define the inelastic resistivity is arbitrary;
resistivity cuts at any other temperature above $T_c$ give qualitatively similar results. 

Now, the intraband inelastic scattering wavevectors that favour $d$-wave pairing in \K~are large-$Q$ processes where $Q$ is the momentum transfer. \cite{thomale_exotic_2011}
By contrast, theoretical calculations show that the $s_{\pm}^h$ pairing state, which changes sign between the $\alpha$ and the $\xi$ hole pockets,
is favoured by small-$Q$ interband interactions. \cite{maiti_gap_2012}
Therefore, one scenario in which to understand the evolution in the inelastic resistivity with pressure (Fig.~\ref{InelasticScattering_universal_Filled}), and its link to the \Tc~reversal, is the following.
At low pressure, the large-$Q$ scattering processes that favor $d$-wave pairing make a substantial contribution to the resistivity, as they produce a large change in momentum.
These weaken with pressure, causing a decrease in both \Tc~and the resistivity.
This decrease persists until the low-$Q$ processes that favor  $s_{\pm}^h$ pairing, less visible in the resistivity, come to dominate, above \Pc. 

\section{\label{Summary}Summary}

In summary, we report a universal V-shaped pressure dependence of \Tc~in AFe$_2$As$_2$ with A = K, Rb, and Cs.
Remarkably, the temperature-pressure superconducting phase diagram of these fully hole-doped iron arsenides is universal,
unaffected by a change in the alkali atom A, with its concomitant changes in structural parameters.
%
%
In the absence of any sudden change in the Fermi surface across the critical pressure \Pc, we interpret the \Tc~reversal at \Pc~as a transition from one pairing state to another.
%
%
The observation of a sudden change in the upper critical field \Hc($T$), which also appears to be universal in that it is identical in \Rb~and \K, 
is compelling evidence of a sudden change in the structure of the superconducting gap across \Pc.
%
Our proposal is a $d$-wave state below \Pc~and a so-called $s_{\pm}^h$ state, where the gap changes sign between $\Gamma$-centered hole pockets, above \Pc.
Our study of the pressure dependence of resistivity in \K, \Cs, and \Rb~reveals a possible link between \Tc~and inelastic scattering.
As the pressure is lowered, the large-$Q$ inelastic scattering processes that favour $d$-wave pairing grow until
at a critical pressure \Pc~they cause a change of pairing symmetry from $s$-wave to $d$-wave.


\section*{ACKNOWLEDGMENTS}

We thank 
J.~Carbotte,
A.~V.~Chubukov, 
J.~P.~Clancy,
R.~M.~Fernandes, 
M.~Franz,
Y.-J.~Kim,
C.~Meingast,
A.~J.~Millis,
R.~Prozorov,
P.~Richard,
M.~A.~Tanatar,
V.~Taufour,
I.~Vekhter,
and
H.~von L\"{o}hneysen
for helpful discussions, as well as
B.~Vincent
and
S.~Fortier 
for assistance with the experiments. 
The work at Sherbrooke was supported by the Canadian Institute for Advanced Research and a Canada Research Chair and it was funded by NSERC,
FRQNT and CFI. 
Work done in China was supported by the National Natural Science Foundation of China (Grant No. 11190021), 
the Strategic Priority Research Program (B) of the Chinese Academy of Sciences, 
and the National Basic Research Program of China.






\bibliography{RbFe2As2_05feb2015}

\begin{thebibliography}{37}%
\makeatletter
\providecommand \@ifxundefined [1]{%
 \@ifx{#1\undefined}
}%
\providecommand \@ifnum [1]{%
 \ifnum #1\expandafter \@firstoftwo
 \else \expandafter \@secondoftwo
 \fi
}%
\providecommand \@ifx [1]{%
 \ifx #1\expandafter \@firstoftwo
 \else \expandafter \@secondoftwo
 \fi
}%
\providecommand \natexlab [1]{#1}%
\providecommand \enquote  [1]{``#1''}%
\providecommand \bibnamefont  [1]{#1}%
\providecommand \bibfnamefont [1]{#1}%
\providecommand \citenamefont [1]{#1}%
\providecommand \href@noop [0]{\@secondoftwo}%
\providecommand \href [0]{\begingroup \@sanitize@url \@href}%
\providecommand \@href[1]{\@@startlink{#1}\@@href}%
\providecommand \@@href[1]{\endgroup#1\@@endlink}%
\providecommand \@sanitize@url [0]{\catcode `\\12\catcode `\$12\catcode
  `\&12\catcode `\#12\catcode `\^12\catcode `\_12\catcode `\%12\relax}%
\providecommand \@@startlink[1]{}%
\providecommand \@@endlink[0]{}%
\providecommand \url  [0]{\begingroup\@sanitize@url \@url }%
\providecommand \@url [1]{\endgroup\@href {#1}{\urlprefix }}%
\providecommand \urlprefix  [0]{URL }%
\providecommand \Eprint [0]{\href }%
\providecommand \doibase [0]{http://dx.doi.org/}%
\providecommand \selectlanguage [0]{\@gobble}%
\providecommand \bibinfo  [0]{\@secondoftwo}%
\providecommand \bibfield  [0]{\@secondoftwo}%
\providecommand \translation [1]{[#1]}%
\providecommand \BibitemOpen [0]{}%
\providecommand \bibitemStop [0]{}%
\providecommand \bibitemNoStop [0]{.\EOS\space}%
\providecommand \EOS [0]{\spacefactor3000\relax}%
\providecommand \BibitemShut  [1]{\csname bibitem#1\endcsname}%
\let\auto@bib@innerbib\@empty
\bibitem [{\citenamefont {Chubukov}(2012)}]{chubukov_pairing_2012}%
  \BibitemOpen
  \bibfield  {author} {\bibinfo {author} {\bibfnamefont {A.}~\bibnamefont
  {Chubukov}},\ }\href {\doibase 10.1146/annurev-conmatphys-020911-125055}
  {\bibfield  {journal} {\bibinfo  {journal} {Annual Review of Condensed Matter
  Physics}\ }\textbf {\bibinfo {volume} {3}},\ \bibinfo {pages} {57} (\bibinfo
  {year} {2012})}\BibitemShut {NoStop}%
\bibitem [{\citenamefont {Hirschfeld}\ \emph {et~al.}(2011)\citenamefont
  {Hirschfeld}, \citenamefont {Korshunov},\ and\ \citenamefont
  {Mazin}}]{hirschfeld_gap_2011}%
  \BibitemOpen
  \bibfield  {author} {\bibinfo {author} {\bibfnamefont {P.~J.}\ \bibnamefont
  {Hirschfeld}}, \bibinfo {author} {\bibfnamefont {M.~M.}\ \bibnamefont
  {Korshunov}}, \ and\ \bibinfo {author} {\bibfnamefont {I.~I.}\ \bibnamefont
  {Mazin}},\ }\href {\doibase 10.1088/0034-4885/74/12/124508} {\bibfield
  {journal} {\bibinfo  {journal} {Reports on Progress in Physics}\ }\textbf
  {\bibinfo {volume} {74}},\ \bibinfo {pages} {124508} (\bibinfo {year}
  {2011})}\BibitemShut {NoStop}%
\bibitem [{\citenamefont {Graser}\ \emph {et~al.}(2009)\citenamefont {Graser},
  \citenamefont {Maier}, \citenamefont {Hirschfeld},\ and\ \citenamefont
  {Scalapino}}]{graser_near-degeneracy_2009}%
  \BibitemOpen
  \bibfield  {author} {\bibinfo {author} {\bibfnamefont {S.}~\bibnamefont
  {Graser}}, \bibinfo {author} {\bibfnamefont {T.~A.}\ \bibnamefont {Maier}},
  \bibinfo {author} {\bibfnamefont {P.~J.}\ \bibnamefont {Hirschfeld}}, \ and\
  \bibinfo {author} {\bibfnamefont {D.~J.}\ \bibnamefont {Scalapino}},\ }\href
  {\doibase 10.1088/1367-2630/11/2/025016} {\bibfield  {journal} {\bibinfo
  {journal} {New Journal of Physics}\ }\textbf {\bibinfo {volume} {11}},\
  \bibinfo {pages} {025016} (\bibinfo {year} {2009})}\BibitemShut {NoStop}%
\bibitem [{\citenamefont {Fernandes}\ and\ \citenamefont
  {Millis}(2013)}]{fernandes_suppression_2013}%
  \BibitemOpen
  \bibfield  {author} {\bibinfo {author} {\bibfnamefont {R.~M.}\ \bibnamefont
  {Fernandes}}\ and\ \bibinfo {author} {\bibfnamefont {A.~J.}\ \bibnamefont
  {Millis}},\ }\href@noop {} {\bibfield  {journal} {\bibinfo  {journal}
  {Physical Review Letters}\ }\textbf {\bibinfo {volume} {110}},\ \bibinfo
  {pages} {117004} (\bibinfo {year} {2013})}\BibitemShut {NoStop}%
\bibitem [{\citenamefont {Tafti}\ \emph {et~al.}(2013)\citenamefont {Tafti},
  \citenamefont {Juneau-Fecteau}, \citenamefont {Delage}, \citenamefont
  {Ren\'{e}~de Cotret}, \citenamefont {Reid}, \citenamefont {Wang},
  \citenamefont {Luo}, \citenamefont {Chen}, \citenamefont {Doiron-Leyraud},\
  and\ \citenamefont {Taillefer}}]{tafti_sudden_2013}%
  \BibitemOpen
  \bibfield  {author} {\bibinfo {author} {\bibfnamefont {F.~F.}\ \bibnamefont
  {Tafti}}, \bibinfo {author} {\bibfnamefont {A.}~\bibnamefont
  {Juneau-Fecteau}}, \bibinfo {author} {\bibfnamefont {M.-E.}\ \bibnamefont
  {Delage}}, \bibinfo {author} {\bibfnamefont {S.}~\bibnamefont {Ren\'{e}~de
  Cotret}}, \bibinfo {author} {\bibfnamefont {J.-P.}\ \bibnamefont {Reid}},
  \bibinfo {author} {\bibfnamefont {A.~F.}\ \bibnamefont {Wang}}, \bibinfo
  {author} {\bibfnamefont {X.-G.}\ \bibnamefont {Luo}}, \bibinfo {author}
  {\bibfnamefont {X.~H.}\ \bibnamefont {Chen}}, \bibinfo {author}
  {\bibfnamefont {N.}~\bibnamefont {Doiron-Leyraud}}, \ and\ \bibinfo {author}
  {\bibfnamefont {L.}~\bibnamefont {Taillefer}},\ }\href {\doibase
  10.1038/nphys2617} {\bibfield  {journal} {\bibinfo  {journal} {Nature
  Physics}\ }\textbf {\bibinfo {volume} {9}},\ \bibinfo {pages} {349} (\bibinfo
  {year} {2013})}\BibitemShut {NoStop}%
\bibitem [{\citenamefont {Tafti}\ \emph {et~al.}(2014)\citenamefont {Tafti},
  \citenamefont {Clancy}, \citenamefont {Lapointe-Major}, \citenamefont
  {Collignon}, \citenamefont {Faucher}, \citenamefont {Sears}, \citenamefont
  {Juneau-Fecteau}, \citenamefont {Doiron-Leyraud}, \citenamefont {Wang},
  \citenamefont {Luo}, \citenamefont {Chen}, \citenamefont {Desgreniers},
  \citenamefont {Kim},\ and\ \citenamefont {Taillefer}}]{tafti_sudden_2014}%
  \BibitemOpen
  \bibfield  {author} {\bibinfo {author} {\bibfnamefont {F.~F.}\ \bibnamefont
  {Tafti}}, \bibinfo {author} {\bibfnamefont {J.~P.}\ \bibnamefont {Clancy}},
  \bibinfo {author} {\bibfnamefont {M.}~\bibnamefont {Lapointe-Major}},
  \bibinfo {author} {\bibfnamefont {C.}~\bibnamefont {Collignon}}, \bibinfo
  {author} {\bibfnamefont {S.}~\bibnamefont {Faucher}}, \bibinfo {author}
  {\bibfnamefont {J.~A.}\ \bibnamefont {Sears}}, \bibinfo {author}
  {\bibfnamefont {A.}~\bibnamefont {Juneau-Fecteau}}, \bibinfo {author}
  {\bibfnamefont {N.}~\bibnamefont {Doiron-Leyraud}}, \bibinfo {author}
  {\bibfnamefont {A.~F.}\ \bibnamefont {Wang}}, \bibinfo {author}
  {\bibfnamefont {X.-G.}\ \bibnamefont {Luo}}, \bibinfo {author} {\bibfnamefont
  {X.~H.}\ \bibnamefont {Chen}}, \bibinfo {author} {\bibfnamefont
  {S.}~\bibnamefont {Desgreniers}}, \bibinfo {author} {\bibfnamefont {Y.-J.}\
  \bibnamefont {Kim}}, \ and\ \bibinfo {author} {\bibfnamefont
  {L.}~\bibnamefont {Taillefer}},\ }\href {\doibase 10.1103/PhysRevB.89.134502}
  {\bibfield  {journal} {\bibinfo  {journal} {Physical Review B}\ }\textbf
  {\bibinfo {volume} {89}},\ \bibinfo {pages} {134502} (\bibinfo {year}
  {2014})}\BibitemShut {NoStop}%
\bibitem [{\citenamefont {Terashima}\ \emph {et~al.}(2014)\citenamefont
  {Terashima}, \citenamefont {Kihou}, \citenamefont {Sugii}, \citenamefont
  {Kikugawa}, \citenamefont {Matsumoto}, \citenamefont {Ishida}, \citenamefont
  {Lee}, \citenamefont {Iyo}, \citenamefont {Eisaki},\ and\ \citenamefont
  {Uji}}]{terashima_two_2014}%
  \BibitemOpen
  \bibfield  {author} {\bibinfo {author} {\bibfnamefont {T.}~\bibnamefont
  {Terashima}}, \bibinfo {author} {\bibfnamefont {K.}~\bibnamefont {Kihou}},
  \bibinfo {author} {\bibfnamefont {K.}~\bibnamefont {Sugii}}, \bibinfo
  {author} {\bibfnamefont {N.}~\bibnamefont {Kikugawa}}, \bibinfo {author}
  {\bibfnamefont {T.}~\bibnamefont {Matsumoto}}, \bibinfo {author}
  {\bibfnamefont {S.}~\bibnamefont {Ishida}}, \bibinfo {author} {\bibfnamefont
  {C.-H.}\ \bibnamefont {Lee}}, \bibinfo {author} {\bibfnamefont
  {A.}~\bibnamefont {Iyo}}, \bibinfo {author} {\bibfnamefont {H.}~\bibnamefont
  {Eisaki}}, \ and\ \bibinfo {author} {\bibfnamefont {S.}~\bibnamefont {Uji}},\
  }\href {\doibase 10.1103/PhysRevB.89.134520} {\bibfield  {journal} {\bibinfo
  {journal} {Physical Review B}\ }\textbf {\bibinfo {volume} {89}},\ \bibinfo
  {pages} {134520} (\bibinfo {year} {2014})}\BibitemShut {NoStop}%
\bibitem [{\citenamefont {Taufour}\ \emph {et~al.}(2014)\citenamefont
  {Taufour}, \citenamefont {Foroozani}, \citenamefont {Tanatar}, \citenamefont
  {Lim}, \citenamefont {Kaluarachchi}, \citenamefont {Kim}, \citenamefont
  {Liu}, \citenamefont {Lograsso}, \citenamefont {Kogan}, \citenamefont
  {Prozorov}, \citenamefont {Bud'ko}, \citenamefont {Schilling},\ and\
  \citenamefont {Canfield}}]{taufour_upper_2014}%
  \BibitemOpen
  \bibfield  {author} {\bibinfo {author} {\bibfnamefont {V.}~\bibnamefont
  {Taufour}}, \bibinfo {author} {\bibfnamefont {N.}~\bibnamefont {Foroozani}},
  \bibinfo {author} {\bibfnamefont {M.~A.}\ \bibnamefont {Tanatar}}, \bibinfo
  {author} {\bibfnamefont {J.}~\bibnamefont {Lim}}, \bibinfo {author}
  {\bibfnamefont {U.}~\bibnamefont {Kaluarachchi}}, \bibinfo {author}
  {\bibfnamefont {S.~K.}\ \bibnamefont {Kim}}, \bibinfo {author} {\bibfnamefont
  {Y.}~\bibnamefont {Liu}}, \bibinfo {author} {\bibfnamefont {T.~A.}\
  \bibnamefont {Lograsso}}, \bibinfo {author} {\bibfnamefont {V.~G.}\
  \bibnamefont {Kogan}}, \bibinfo {author} {\bibfnamefont {R.}~\bibnamefont
  {Prozorov}}, \bibinfo {author} {\bibfnamefont {S.~L.}\ \bibnamefont
  {Bud'ko}}, \bibinfo {author} {\bibfnamefont {J.~S.}\ \bibnamefont
  {Schilling}}, \ and\ \bibinfo {author} {\bibfnamefont {P.~C.}\ \bibnamefont
  {Canfield}},\ }\href {\doibase 10.1103/PhysRevB.89.220509} {\bibfield
  {journal} {\bibinfo  {journal} {Physical Review B}\ }\textbf {\bibinfo
  {volume} {89}},\ \bibinfo {pages} {220509} (\bibinfo {year}
  {2014})}\BibitemShut {NoStop}%
\bibitem [{\citenamefont {Grinenko}\ \emph
  {et~al.}(2014{\natexlab{a}})\citenamefont {Grinenko}, \citenamefont
  {Schottenhamel}, \citenamefont {Wolter}, \citenamefont {Efremov},
  \citenamefont {Drechsler}, \citenamefont {Aswartham}, \citenamefont {Kumar},
  \citenamefont {Wurmehl}, \citenamefont {Roslova}, \citenamefont {Morozov},
  \citenamefont {Holzapfel}, \citenamefont {B\"{u}chner}, \citenamefont
  {Ahrens}, \citenamefont {Troyanov}, \citenamefont {K\"{o}hler}, \citenamefont
  {Gati}, \citenamefont {Kn\"{o}ner}, \citenamefont {Hoang}, \citenamefont
  {Lang}, \citenamefont {Ricci},\ and\ \citenamefont
  {Profeta}}]{grinenko_superconducting_2014}%
  \BibitemOpen
  \bibfield  {author} {\bibinfo {author} {\bibfnamefont {V.}~\bibnamefont
  {Grinenko}}, \bibinfo {author} {\bibfnamefont {W.}~\bibnamefont
  {Schottenhamel}}, \bibinfo {author} {\bibfnamefont {A.~U.~B.}\ \bibnamefont
  {Wolter}}, \bibinfo {author} {\bibfnamefont {D.~V.}\ \bibnamefont {Efremov}},
  \bibinfo {author} {\bibfnamefont {S.-L.}\ \bibnamefont {Drechsler}}, \bibinfo
  {author} {\bibfnamefont {S.}~\bibnamefont {Aswartham}}, \bibinfo {author}
  {\bibfnamefont {M.}~\bibnamefont {Kumar}}, \bibinfo {author} {\bibfnamefont
  {S.}~\bibnamefont {Wurmehl}}, \bibinfo {author} {\bibfnamefont
  {M.}~\bibnamefont {Roslova}}, \bibinfo {author} {\bibfnamefont {I.~V.}\
  \bibnamefont {Morozov}}, \bibinfo {author} {\bibfnamefont {B.}~\bibnamefont
  {Holzapfel}}, \bibinfo {author} {\bibfnamefont {B.}~\bibnamefont
  {B\"{u}chner}}, \bibinfo {author} {\bibfnamefont {E.}~\bibnamefont {Ahrens}},
  \bibinfo {author} {\bibfnamefont {S.~I.}\ \bibnamefont {Troyanov}}, \bibinfo
  {author} {\bibfnamefont {S.}~\bibnamefont {K\"{o}hler}}, \bibinfo {author}
  {\bibfnamefont {E.}~\bibnamefont {Gati}}, \bibinfo {author} {\bibfnamefont
  {S.}~\bibnamefont {Kn\"{o}ner}}, \bibinfo {author} {\bibfnamefont {N.~H.}\
  \bibnamefont {Hoang}}, \bibinfo {author} {\bibfnamefont {M.}~\bibnamefont
  {Lang}}, \bibinfo {author} {\bibfnamefont {F.}~\bibnamefont {Ricci}}, \ and\
  \bibinfo {author} {\bibfnamefont {G.}~\bibnamefont {Profeta}},\ }\href
  {\doibase 10.1103/PhysRevB.90.094511} {\bibfield  {journal} {\bibinfo
  {journal} {Physical Review B}\ }\textbf {\bibinfo {volume} {90}},\ \bibinfo
  {pages} {094511} (\bibinfo {year} {2014}{\natexlab{a}})}\BibitemShut
  {NoStop}%
\bibitem [{\citenamefont {Terashima}\ \emph {et~al.}(2010)\citenamefont
  {Terashima}, \citenamefont {Kimata}, \citenamefont {Kurita}, \citenamefont
  {Satsukawa}, \citenamefont {Harada}, \citenamefont {Hazama}, \citenamefont
  {Imai}, \citenamefont {Sato}, \citenamefont {Kihou}, \citenamefont {Lee},
  \citenamefont {Kito}, \citenamefont {Eisaki}, \citenamefont {Iyo},
  \citenamefont {Saito}, \citenamefont {Fukazawa}, \citenamefont {Kohori},
  \citenamefont {Harima},\ and\ \citenamefont {Uji}}]{terashima_fermi_2010}%
  \BibitemOpen
  \bibfield  {author} {\bibinfo {author} {\bibfnamefont {T.}~\bibnamefont
  {Terashima}}, \bibinfo {author} {\bibfnamefont {M.}~\bibnamefont {Kimata}},
  \bibinfo {author} {\bibfnamefont {N.}~\bibnamefont {Kurita}}, \bibinfo
  {author} {\bibfnamefont {H.}~\bibnamefont {Satsukawa}}, \bibinfo {author}
  {\bibfnamefont {A.}~\bibnamefont {Harada}}, \bibinfo {author} {\bibfnamefont
  {K.}~\bibnamefont {Hazama}}, \bibinfo {author} {\bibfnamefont
  {M.}~\bibnamefont {Imai}}, \bibinfo {author} {\bibfnamefont {A.}~\bibnamefont
  {Sato}}, \bibinfo {author} {\bibfnamefont {K.}~\bibnamefont {Kihou}},
  \bibinfo {author} {\bibfnamefont {C.-H.}\ \bibnamefont {Lee}}, \bibinfo
  {author} {\bibfnamefont {H.}~\bibnamefont {Kito}}, \bibinfo {author}
  {\bibfnamefont {H.}~\bibnamefont {Eisaki}}, \bibinfo {author} {\bibfnamefont
  {A.}~\bibnamefont {Iyo}}, \bibinfo {author} {\bibfnamefont {T.}~\bibnamefont
  {Saito}}, \bibinfo {author} {\bibfnamefont {H.}~\bibnamefont {Fukazawa}},
  \bibinfo {author} {\bibfnamefont {Y.}~\bibnamefont {Kohori}}, \bibinfo
  {author} {\bibfnamefont {H.}~\bibnamefont {Harima}}, \ and\ \bibinfo {author}
  {\bibfnamefont {S.}~\bibnamefont {Uji}},\ }\href {\doibase
  10.1143/JPSJ.79.053702} {\bibfield  {journal} {\bibinfo  {journal} {Journal
  of the Physical Society of Japan}\ }\textbf {\bibinfo {volume} {79}},\
  \bibinfo {pages} {053702} (\bibinfo {year} {2010})}\BibitemShut {NoStop}%
\bibitem [{\citenamefont {Terashima}\ \emph {et~al.}(2013)\citenamefont
  {Terashima}, \citenamefont {Kurita}, \citenamefont {Kimata}, \citenamefont
  {Tomita}, \citenamefont {Tsuchiya}, \citenamefont {Imai}, \citenamefont
  {Sato}, \citenamefont {Kihou}, \citenamefont {Lee}, \citenamefont {Kito},
  \citenamefont {Eisaki}, \citenamefont {Iyo}, \citenamefont {Saito},
  \citenamefont {Fukazawa}, \citenamefont {Kohori}, \citenamefont {Harima},\
  and\ \citenamefont {Uji}}]{terashima_fermi_2013}%
  \BibitemOpen
  \bibfield  {author} {\bibinfo {author} {\bibfnamefont {T.}~\bibnamefont
  {Terashima}}, \bibinfo {author} {\bibfnamefont {N.}~\bibnamefont {Kurita}},
  \bibinfo {author} {\bibfnamefont {M.}~\bibnamefont {Kimata}}, \bibinfo
  {author} {\bibfnamefont {M.}~\bibnamefont {Tomita}}, \bibinfo {author}
  {\bibfnamefont {S.}~\bibnamefont {Tsuchiya}}, \bibinfo {author}
  {\bibfnamefont {M.}~\bibnamefont {Imai}}, \bibinfo {author} {\bibfnamefont
  {A.}~\bibnamefont {Sato}}, \bibinfo {author} {\bibfnamefont {K.}~\bibnamefont
  {Kihou}}, \bibinfo {author} {\bibfnamefont {C.-H.}\ \bibnamefont {Lee}},
  \bibinfo {author} {\bibfnamefont {H.}~\bibnamefont {Kito}}, \bibinfo {author}
  {\bibfnamefont {H.}~\bibnamefont {Eisaki}}, \bibinfo {author} {\bibfnamefont
  {A.}~\bibnamefont {Iyo}}, \bibinfo {author} {\bibfnamefont {T.}~\bibnamefont
  {Saito}}, \bibinfo {author} {\bibfnamefont {H.}~\bibnamefont {Fukazawa}},
  \bibinfo {author} {\bibfnamefont {Y.}~\bibnamefont {Kohori}}, \bibinfo
  {author} {\bibfnamefont {H.}~\bibnamefont {Harima}}, \ and\ \bibinfo {author}
  {\bibfnamefont {S.}~\bibnamefont {Uji}},\ }\href@noop {} {\bibfield
  {journal} {\bibinfo  {journal} {Physical Review B}\ }\textbf {\bibinfo
  {volume} {87}},\ \bibinfo {pages} {224512} (\bibinfo {year}
  {2013})}\BibitemShut {NoStop}%
\bibitem [{\citenamefont {{Diego A. Zocco}}\ \emph {et~al.}(2014)\citenamefont
  {{Diego A. Zocco}}, \citenamefont {{Kai Grube}}, \citenamefont {{Felix
  Eilers}}, \citenamefont {{Thomas Wolf}},\ and\ \citenamefont {{Hilbert v.
  L\"{o}hneysen}}}]{diego_a._zocco_fermi_2014}%
  \BibitemOpen
  \bibfield  {author} {\bibinfo {author} {\bibnamefont {{Diego A. Zocco}}},
  \bibinfo {author} {\bibnamefont {{Kai Grube}}}, \bibinfo {author}
  {\bibnamefont {{Felix Eilers}}}, \bibinfo {author} {\bibnamefont {{Thomas
  Wolf}}}, \ and\ \bibinfo {author} {\bibnamefont {{Hilbert v.
  L\"{o}hneysen}}},\ }in\ \href
  {http://journals.jps.jp/doi/abs/10.7566/JPSCP.3.015007} {\emph {\bibinfo
  {booktitle} {Proceedings of the International Conference on Strongly
  Correlated Electron Systems}}},\ \bibinfo {series} {{JPS} Conference
  Proceedings}, Vol.~\bibinfo {volume} {3}\ (\bibinfo  {publisher} {Journal of
  the Physical Society of Japan},\ \bibinfo {year} {2014})\ p.\ \bibinfo
  {pages} {015007}\BibitemShut {NoStop}%
\bibitem [{\citenamefont {Okazaki}\ \emph {et~al.}(2012)\citenamefont
  {Okazaki}, \citenamefont {Ota}, \citenamefont {Kotani}, \citenamefont
  {Malaeb}, \citenamefont {Ishida}, \citenamefont {Shimojima}, \citenamefont
  {Kiss}, \citenamefont {Watanabe}, \citenamefont {Chen}, \citenamefont
  {Kihou}, \citenamefont {Lee}, \citenamefont {Iyo}, \citenamefont {Eisaki},
  \citenamefont {Saito}, \citenamefont {Fukazawa}, \citenamefont {Kohori},
  \citenamefont {Hashimoto}, \citenamefont {Shibauchi}, \citenamefont
  {Matsuda}, \citenamefont {Ikeda}, \citenamefont {Miyahara}, \citenamefont
  {Arita}, \citenamefont {Chainani},\ and\ \citenamefont
  {Shin}}]{okazaki_octet-line_2012}%
  \BibitemOpen
  \bibfield  {author} {\bibinfo {author} {\bibfnamefont {K.}~\bibnamefont
  {Okazaki}}, \bibinfo {author} {\bibfnamefont {Y.}~\bibnamefont {Ota}},
  \bibinfo {author} {\bibfnamefont {Y.}~\bibnamefont {Kotani}}, \bibinfo
  {author} {\bibfnamefont {W.}~\bibnamefont {Malaeb}}, \bibinfo {author}
  {\bibfnamefont {Y.}~\bibnamefont {Ishida}}, \bibinfo {author} {\bibfnamefont
  {T.}~\bibnamefont {Shimojima}}, \bibinfo {author} {\bibfnamefont
  {T.}~\bibnamefont {Kiss}}, \bibinfo {author} {\bibfnamefont {S.}~\bibnamefont
  {Watanabe}}, \bibinfo {author} {\bibfnamefont {C.-T.}\ \bibnamefont {Chen}},
  \bibinfo {author} {\bibfnamefont {K.}~\bibnamefont {Kihou}}, \bibinfo
  {author} {\bibfnamefont {C.~H.}\ \bibnamefont {Lee}}, \bibinfo {author}
  {\bibfnamefont {A.}~\bibnamefont {Iyo}}, \bibinfo {author} {\bibfnamefont
  {H.}~\bibnamefont {Eisaki}}, \bibinfo {author} {\bibfnamefont
  {T.}~\bibnamefont {Saito}}, \bibinfo {author} {\bibfnamefont
  {H.}~\bibnamefont {Fukazawa}}, \bibinfo {author} {\bibfnamefont
  {Y.}~\bibnamefont {Kohori}}, \bibinfo {author} {\bibfnamefont
  {K.}~\bibnamefont {Hashimoto}}, \bibinfo {author} {\bibfnamefont
  {T.}~\bibnamefont {Shibauchi}}, \bibinfo {author} {\bibfnamefont
  {Y.}~\bibnamefont {Matsuda}}, \bibinfo {author} {\bibfnamefont
  {H.}~\bibnamefont {Ikeda}}, \bibinfo {author} {\bibfnamefont
  {H.}~\bibnamefont {Miyahara}}, \bibinfo {author} {\bibfnamefont
  {R.}~\bibnamefont {Arita}}, \bibinfo {author} {\bibfnamefont
  {A.}~\bibnamefont {Chainani}}, \ and\ \bibinfo {author} {\bibfnamefont
  {S.}~\bibnamefont {Shin}},\ }\href {zotero://attachment/360/} {\bibfield
  {journal} {\bibinfo  {journal} {Science}\ }\textbf {\bibinfo {volume}
  {337}},\ \bibinfo {pages} {1314} (\bibinfo {year} {2012})}\BibitemShut
  {NoStop}%
\bibitem [{\citenamefont {Watanabe}\ \emph {et~al.}(2014)\citenamefont
  {Watanabe}, \citenamefont {Yamashita}, \citenamefont {Kawamoto},
  \citenamefont {Kurata}, \citenamefont {Mizukami}, \citenamefont {Ohta},
  \citenamefont {Kasahara}, \citenamefont {Yamashita}, \citenamefont {Saito},
  \citenamefont {Fukazawa}, \citenamefont {Kohori}, \citenamefont {Ishida},
  \citenamefont {Kihou}, \citenamefont {Lee}, \citenamefont {Iyo},
  \citenamefont {Eisaki}, \citenamefont {Vorontsov}, \citenamefont
  {Shibauchi},\ and\ \citenamefont {Matsuda}}]{watanabe_doping_2014}%
  \BibitemOpen
  \bibfield  {author} {\bibinfo {author} {\bibfnamefont {D.}~\bibnamefont
  {Watanabe}}, \bibinfo {author} {\bibfnamefont {T.}~\bibnamefont {Yamashita}},
  \bibinfo {author} {\bibfnamefont {Y.}~\bibnamefont {Kawamoto}}, \bibinfo
  {author} {\bibfnamefont {S.}~\bibnamefont {Kurata}}, \bibinfo {author}
  {\bibfnamefont {Y.}~\bibnamefont {Mizukami}}, \bibinfo {author}
  {\bibfnamefont {T.}~\bibnamefont {Ohta}}, \bibinfo {author} {\bibfnamefont
  {S.}~\bibnamefont {Kasahara}}, \bibinfo {author} {\bibfnamefont
  {M.}~\bibnamefont {Yamashita}}, \bibinfo {author} {\bibfnamefont
  {T.}~\bibnamefont {Saito}}, \bibinfo {author} {\bibfnamefont
  {H.}~\bibnamefont {Fukazawa}}, \bibinfo {author} {\bibfnamefont
  {Y.}~\bibnamefont {Kohori}}, \bibinfo {author} {\bibfnamefont
  {S.}~\bibnamefont {Ishida}}, \bibinfo {author} {\bibfnamefont
  {K.}~\bibnamefont {Kihou}}, \bibinfo {author} {\bibfnamefont {C.~H.}\
  \bibnamefont {Lee}}, \bibinfo {author} {\bibfnamefont {A.}~\bibnamefont
  {Iyo}}, \bibinfo {author} {\bibfnamefont {H.}~\bibnamefont {Eisaki}},
  \bibinfo {author} {\bibfnamefont {A.~B.}\ \bibnamefont {Vorontsov}}, \bibinfo
  {author} {\bibfnamefont {T.}~\bibnamefont {Shibauchi}}, \ and\ \bibinfo
  {author} {\bibfnamefont {Y.}~\bibnamefont {Matsuda}},\ }\href {\doibase
  10.1103/PhysRevB.89.115112} {\bibfield  {journal} {\bibinfo  {journal}
  {Physical Review B}\ }\textbf {\bibinfo {volume} {89}},\ \bibinfo {pages}
  {115112} (\bibinfo {year} {2014})}\BibitemShut {NoStop}%
\bibitem [{\citenamefont {Reid}\ \emph
  {et~al.}(2012{\natexlab{a}})\citenamefont {Reid}, \citenamefont {Tanatar},
  \citenamefont {Juneau-Fecteau}, \citenamefont {Gordon}, \citenamefont
  {Ren\'{e}~de Cotret}, \citenamefont {Doiron-Leyraud}, \citenamefont {Seito},
  \citenamefont {Fukazawa}, \citenamefont {Kohori}, \citenamefont {Kihou},
  \citenamefont {Lee}, \citenamefont {Iyo}, \citenamefont {Eisaki},
  \citenamefont {Prozorov},\ and\ \citenamefont
  {Taillefer}}]{reid_universal_2012}%
  \BibitemOpen
  \bibfield  {author} {\bibinfo {author} {\bibfnamefont {J.-P.}\ \bibnamefont
  {Reid}}, \bibinfo {author} {\bibfnamefont {M.~A.}\ \bibnamefont {Tanatar}},
  \bibinfo {author} {\bibfnamefont {A.}~\bibnamefont {Juneau-Fecteau}},
  \bibinfo {author} {\bibfnamefont {R.~T.}\ \bibnamefont {Gordon}}, \bibinfo
  {author} {\bibfnamefont {S.}~\bibnamefont {Ren\'{e}~de Cotret}}, \bibinfo
  {author} {\bibfnamefont {N.}~\bibnamefont {Doiron-Leyraud}}, \bibinfo
  {author} {\bibfnamefont {T.}~\bibnamefont {Seito}}, \bibinfo {author}
  {\bibfnamefont {H.}~\bibnamefont {Fukazawa}}, \bibinfo {author}
  {\bibfnamefont {Y.}~\bibnamefont {Kohori}}, \bibinfo {author} {\bibfnamefont
  {K.}~\bibnamefont {Kihou}}, \bibinfo {author} {\bibfnamefont {C.~H.}\
  \bibnamefont {Lee}}, \bibinfo {author} {\bibfnamefont {A.}~\bibnamefont
  {Iyo}}, \bibinfo {author} {\bibfnamefont {H.}~\bibnamefont {Eisaki}},
  \bibinfo {author} {\bibfnamefont {R.}~\bibnamefont {Prozorov}}, \ and\
  \bibinfo {author} {\bibfnamefont {L.}~\bibnamefont {Taillefer}},\ }\href@noop
  {} {\bibfield  {journal} {\bibinfo  {journal} {Physical Review Letters}\
  }\textbf {\bibinfo {volume} {109}},\ \bibinfo {pages} {087001} (\bibinfo
  {year} {2012}{\natexlab{a}})}\BibitemShut {NoStop}%
\bibitem [{\citenamefont {Reid}\ \emph
  {et~al.}(2012{\natexlab{b}})\citenamefont {Reid}, \citenamefont
  {Juneau-Fecteau}, \citenamefont {Gordon}, \citenamefont {Ren\'{e}~de Cotret},
  \citenamefont {Doiron-Leyraud}, \citenamefont {Luo}, \citenamefont
  {Shakeripour}, \citenamefont {Chang}, \citenamefont {Tanatar}, \citenamefont
  {Kim}, \citenamefont {Prozorov}, \citenamefont {Saito}, \citenamefont
  {Fukazawa}, \citenamefont {Kohori}, \citenamefont {Kihou}, \citenamefont
  {Lee}, \citenamefont {Iyo}, \citenamefont {Eisaki}, \citenamefont {Shen},
  \citenamefont {Wen},\ and\ \citenamefont {Taillefer}}]{reid_d-wave_2012}%
  \BibitemOpen
  \bibfield  {author} {\bibinfo {author} {\bibfnamefont {J.-P.}\ \bibnamefont
  {Reid}}, \bibinfo {author} {\bibfnamefont {A.}~\bibnamefont
  {Juneau-Fecteau}}, \bibinfo {author} {\bibfnamefont {R.~T.}\ \bibnamefont
  {Gordon}}, \bibinfo {author} {\bibfnamefont {S.}~\bibnamefont {Ren\'{e}~de
  Cotret}}, \bibinfo {author} {\bibfnamefont {N.}~\bibnamefont
  {Doiron-Leyraud}}, \bibinfo {author} {\bibfnamefont {X.-G.}\ \bibnamefont
  {Luo}}, \bibinfo {author} {\bibfnamefont {H.}~\bibnamefont {Shakeripour}},
  \bibinfo {author} {\bibfnamefont {J.}~\bibnamefont {Chang}}, \bibinfo
  {author} {\bibfnamefont {M.~A.}\ \bibnamefont {Tanatar}}, \bibinfo {author}
  {\bibfnamefont {H.}~\bibnamefont {Kim}}, \bibinfo {author} {\bibfnamefont
  {R.}~\bibnamefont {Prozorov}}, \bibinfo {author} {\bibfnamefont
  {T.}~\bibnamefont {Saito}}, \bibinfo {author} {\bibfnamefont
  {H.}~\bibnamefont {Fukazawa}}, \bibinfo {author} {\bibfnamefont
  {Y.}~\bibnamefont {Kohori}}, \bibinfo {author} {\bibfnamefont
  {K.}~\bibnamefont {Kihou}}, \bibinfo {author} {\bibfnamefont {C.~H.}\
  \bibnamefont {Lee}}, \bibinfo {author} {\bibfnamefont {A.}~\bibnamefont
  {Iyo}}, \bibinfo {author} {\bibfnamefont {H.}~\bibnamefont {Eisaki}},
  \bibinfo {author} {\bibfnamefont {B.}~\bibnamefont {Shen}}, \bibinfo {author}
  {\bibfnamefont {H.-H.}\ \bibnamefont {Wen}}, \ and\ \bibinfo {author}
  {\bibfnamefont {L.}~\bibnamefont {Taillefer}},\ }\href
  {zotero://attachment/735/} {\bibfield  {journal} {\bibinfo  {journal}
  {Superconductor Science and Technology}\ }\textbf {\bibinfo {volume} {25}},\
  \bibinfo {pages} {084013} (\bibinfo {year} {2012}{\natexlab{b}})}\BibitemShut
  {NoStop}%
\bibitem [{\citenamefont {Abdel-Hafiez}\ \emph {et~al.}(2013)\citenamefont
  {Abdel-Hafiez}, \citenamefont {Grinenko}, \citenamefont {Aswartham},
  \citenamefont {Morozov}, \citenamefont {Roslova}, \citenamefont {Vakaliuk},
  \citenamefont {Johnston}, \citenamefont {Efremov}, \citenamefont {van~den
  Brink}, \citenamefont {Rosner}, \citenamefont {Kumar}, \citenamefont {Hess},
  \citenamefont {Wurmehl}, \citenamefont {Wolter}, \citenamefont {B\"{u}chner},
  \citenamefont {Green}, \citenamefont {Wosnitza}, \citenamefont {Vogt},
  \citenamefont {Reifenberger}, \citenamefont {Enss}, \citenamefont {Hempel},
  \citenamefont {Klingeler},\ and\ \citenamefont
  {Drechsler}}]{abdel-hafiez_evidence_2013}%
  \BibitemOpen
  \bibfield  {author} {\bibinfo {author} {\bibfnamefont {M.}~\bibnamefont
  {Abdel-Hafiez}}, \bibinfo {author} {\bibfnamefont {V.}~\bibnamefont
  {Grinenko}}, \bibinfo {author} {\bibfnamefont {S.}~\bibnamefont {Aswartham}},
  \bibinfo {author} {\bibfnamefont {I.}~\bibnamefont {Morozov}}, \bibinfo
  {author} {\bibfnamefont {M.}~\bibnamefont {Roslova}}, \bibinfo {author}
  {\bibfnamefont {O.}~\bibnamefont {Vakaliuk}}, \bibinfo {author}
  {\bibfnamefont {S.}~\bibnamefont {Johnston}}, \bibinfo {author}
  {\bibfnamefont {D.~V.}\ \bibnamefont {Efremov}}, \bibinfo {author}
  {\bibfnamefont {J.}~\bibnamefont {van~den Brink}}, \bibinfo {author}
  {\bibfnamefont {H.}~\bibnamefont {Rosner}}, \bibinfo {author} {\bibfnamefont
  {M.}~\bibnamefont {Kumar}}, \bibinfo {author} {\bibfnamefont
  {C.}~\bibnamefont {Hess}}, \bibinfo {author} {\bibfnamefont {S.}~\bibnamefont
  {Wurmehl}}, \bibinfo {author} {\bibfnamefont {A.~U.~B.}\ \bibnamefont
  {Wolter}}, \bibinfo {author} {\bibfnamefont {B.}~\bibnamefont {B\"{u}chner}},
  \bibinfo {author} {\bibfnamefont {E.~L.}\ \bibnamefont {Green}}, \bibinfo
  {author} {\bibfnamefont {J.}~\bibnamefont {Wosnitza}}, \bibinfo {author}
  {\bibfnamefont {P.}~\bibnamefont {Vogt}}, \bibinfo {author} {\bibfnamefont
  {A.}~\bibnamefont {Reifenberger}}, \bibinfo {author} {\bibfnamefont
  {C.}~\bibnamefont {Enss}}, \bibinfo {author} {\bibfnamefont {M.}~\bibnamefont
  {Hempel}}, \bibinfo {author} {\bibfnamefont {R.}~\bibnamefont {Klingeler}}, \
  and\ \bibinfo {author} {\bibfnamefont {S.-L.}\ \bibnamefont {Drechsler}},\
  }\href {\doibase 10.1103/PhysRevB.87.180507} {\bibfield  {journal} {\bibinfo
  {journal} {Physical Review B}\ }\textbf {\bibinfo {volume} {87}},\ \bibinfo
  {pages} {180507} (\bibinfo {year} {2013})}\BibitemShut {NoStop}%
\bibitem [{\citenamefont {Grinenko}\ \emph
  {et~al.}(2014{\natexlab{b}})\citenamefont {Grinenko}, \citenamefont
  {Efremov}, \citenamefont {Drechsler}, \citenamefont {Aswartham},
  \citenamefont {Gruner}, \citenamefont {Roslova}, \citenamefont {Morozov},
  \citenamefont {Nenkov}, \citenamefont {Wurmehl}, \citenamefont {Wolter},
  \citenamefont {Holzapfel},\ and\ \citenamefont
  {B\"{u}chner}}]{grinenko_superconducting_Cv_2014}%
  \BibitemOpen
  \bibfield  {author} {\bibinfo {author} {\bibfnamefont {V.}~\bibnamefont
  {Grinenko}}, \bibinfo {author} {\bibfnamefont {D.~V.}\ \bibnamefont
  {Efremov}}, \bibinfo {author} {\bibfnamefont {S.-L.}\ \bibnamefont
  {Drechsler}}, \bibinfo {author} {\bibfnamefont {S.}~\bibnamefont
  {Aswartham}}, \bibinfo {author} {\bibfnamefont {D.}~\bibnamefont {Gruner}},
  \bibinfo {author} {\bibfnamefont {M.}~\bibnamefont {Roslova}}, \bibinfo
  {author} {\bibfnamefont {I.}~\bibnamefont {Morozov}}, \bibinfo {author}
  {\bibfnamefont {K.}~\bibnamefont {Nenkov}}, \bibinfo {author} {\bibfnamefont
  {S.}~\bibnamefont {Wurmehl}}, \bibinfo {author} {\bibfnamefont {A.~U.~B.}\
  \bibnamefont {Wolter}}, \bibinfo {author} {\bibfnamefont {B.}~\bibnamefont
  {Holzapfel}}, \ and\ \bibinfo {author} {\bibfnamefont {B.}~\bibnamefont
  {B\"{u}chner}},\ }\href {\doibase 10.1103/PhysRevB.89.060504} {\bibfield
  {journal} {\bibinfo  {journal} {Physical Review B}\ }\textbf {\bibinfo
  {volume} {89}},\ \bibinfo {pages} {060504} (\bibinfo {year}
  {2014}{\natexlab{b}})}\BibitemShut {NoStop}%
\bibitem [{\citenamefont {Hashimoto}\ \emph {et~al.}(2010)\citenamefont
  {Hashimoto}, \citenamefont {Serafin}, \citenamefont {Tonegawa}, \citenamefont
  {Katsumata}, \citenamefont {Okazaki}, \citenamefont {Saito}, \citenamefont
  {Fukazawa}, \citenamefont {Kohori}, \citenamefont {Kihou}, \citenamefont
  {Lee}, \citenamefont {Iyo}, \citenamefont {Eisaki}, \citenamefont {Ikeda},
  \citenamefont {Matsuda}, \citenamefont {Carrington},\ and\ \citenamefont
  {Shibauchi}}]{hashimoto_evidence_2010}%
  \BibitemOpen
  \bibfield  {author} {\bibinfo {author} {\bibfnamefont {K.}~\bibnamefont
  {Hashimoto}}, \bibinfo {author} {\bibfnamefont {A.}~\bibnamefont {Serafin}},
  \bibinfo {author} {\bibfnamefont {S.}~\bibnamefont {Tonegawa}}, \bibinfo
  {author} {\bibfnamefont {R.}~\bibnamefont {Katsumata}}, \bibinfo {author}
  {\bibfnamefont {R.}~\bibnamefont {Okazaki}}, \bibinfo {author} {\bibfnamefont
  {T.}~\bibnamefont {Saito}}, \bibinfo {author} {\bibfnamefont
  {H.}~\bibnamefont {Fukazawa}}, \bibinfo {author} {\bibfnamefont
  {Y.}~\bibnamefont {Kohori}}, \bibinfo {author} {\bibfnamefont
  {K.}~\bibnamefont {Kihou}}, \bibinfo {author} {\bibfnamefont {C.~H.}\
  \bibnamefont {Lee}}, \bibinfo {author} {\bibfnamefont {A.}~\bibnamefont
  {Iyo}}, \bibinfo {author} {\bibfnamefont {H.}~\bibnamefont {Eisaki}},
  \bibinfo {author} {\bibfnamefont {H.}~\bibnamefont {Ikeda}}, \bibinfo
  {author} {\bibfnamefont {Y.}~\bibnamefont {Matsuda}}, \bibinfo {author}
  {\bibfnamefont {A.}~\bibnamefont {Carrington}}, \ and\ \bibinfo {author}
  {\bibfnamefont {T.}~\bibnamefont {Shibauchi}},\ }\href {\doibase
  10.1103/PhysRevB.82.014526} {\bibfield  {journal} {\bibinfo  {journal}
  {Physical Review B}\ }\textbf {\bibinfo {volume} {82}},\ \bibinfo {pages}
  {014526} (\bibinfo {year} {2010})}\BibitemShut {NoStop}%
\bibitem [{\citenamefont {Kim}\ \emph {et~al.}(2014)\citenamefont {Kim},
  \citenamefont {Tanatar}, \citenamefont {Liu}, \citenamefont {Sims},
  \citenamefont {Zhang}, \citenamefont {Dai}, \citenamefont {Lograsso},\ and\
  \citenamefont {Prozorov}}]{kim_evolution_2014}%
  \BibitemOpen
  \bibfield  {author} {\bibinfo {author} {\bibfnamefont {H.}~\bibnamefont
  {Kim}}, \bibinfo {author} {\bibfnamefont {M.~A.}\ \bibnamefont {Tanatar}},
  \bibinfo {author} {\bibfnamefont {Y.}~\bibnamefont {Liu}}, \bibinfo {author}
  {\bibfnamefont {Z.~C.}\ \bibnamefont {Sims}}, \bibinfo {author}
  {\bibfnamefont {C.}~\bibnamefont {Zhang}}, \bibinfo {author} {\bibfnamefont
  {P.}~\bibnamefont {Dai}}, \bibinfo {author} {\bibfnamefont {T.~A.}\
  \bibnamefont {Lograsso}}, \ and\ \bibinfo {author} {\bibfnamefont
  {R.}~\bibnamefont {Prozorov}},\ }\href {\doibase 10.1103/PhysRevB.89.174519}
  {\bibfield  {journal} {\bibinfo  {journal} {Physical Review B}\ }\textbf
  {\bibinfo {volume} {89}},\ \bibinfo {pages} {174519} (\bibinfo {year}
  {2014})}\BibitemShut {NoStop}%
\bibitem [{\citenamefont {Wang}\ \emph {et~al.}(2013)\citenamefont {Wang},
  \citenamefont {Pan}, \citenamefont {Luo}, \citenamefont {Chen}, \citenamefont
  {Yan}, \citenamefont {Ying}, \citenamefont {Ye}, \citenamefont {Cheng},
  \citenamefont {Hong}, \citenamefont {Li},\ and\ \citenamefont
  {Chen}}]{wang_calorimetric_2013}%
  \BibitemOpen
  \bibfield  {author} {\bibinfo {author} {\bibfnamefont {A.~F.}\ \bibnamefont
  {Wang}}, \bibinfo {author} {\bibfnamefont {B.~Y.}\ \bibnamefont {Pan}},
  \bibinfo {author} {\bibfnamefont {X.~G.}\ \bibnamefont {Luo}}, \bibinfo
  {author} {\bibfnamefont {F.}~\bibnamefont {Chen}}, \bibinfo {author}
  {\bibfnamefont {Y.~J.}\ \bibnamefont {Yan}}, \bibinfo {author} {\bibfnamefont
  {J.~J.}\ \bibnamefont {Ying}}, \bibinfo {author} {\bibfnamefont {G.~J.}\
  \bibnamefont {Ye}}, \bibinfo {author} {\bibfnamefont {P.}~\bibnamefont
  {Cheng}}, \bibinfo {author} {\bibfnamefont {X.~C.}\ \bibnamefont {Hong}},
  \bibinfo {author} {\bibfnamefont {S.~Y.}\ \bibnamefont {Li}}, \ and\ \bibinfo
  {author} {\bibfnamefont {X.~H.}\ \bibnamefont {Chen}},\ }\href {\doibase
  10.1103/PhysRevB.87.214509} {\bibfield  {journal} {\bibinfo  {journal}
  {Physical Review B}\ }\textbf {\bibinfo {volume} {87}},\ \bibinfo {pages}
  {214509} (\bibinfo {year} {2013})}\BibitemShut {NoStop}%
\bibitem [{\citenamefont {Zhang}\ \emph {et~al.}(2014)\citenamefont {Zhang},
  \citenamefont {Wang}, \citenamefont {Hong}, \citenamefont {Zhang},
  \citenamefont {Pan}, \citenamefont {Pan}, \citenamefont {Xu}, \citenamefont
  {Luo}, \citenamefont {Chen},\ and\ \citenamefont {Li}}]{zhang_heat_2014}%
  \BibitemOpen
  \bibfield  {author} {\bibinfo {author} {\bibfnamefont {Z.}~\bibnamefont
  {Zhang}}, \bibinfo {author} {\bibfnamefont {A.~F.}\ \bibnamefont {Wang}},
  \bibinfo {author} {\bibfnamefont {X.~C.}\ \bibnamefont {Hong}}, \bibinfo
  {author} {\bibfnamefont {J.}~\bibnamefont {Zhang}}, \bibinfo {author}
  {\bibfnamefont {B.~Y.}\ \bibnamefont {Pan}}, \bibinfo {author} {\bibfnamefont
  {J.}~\bibnamefont {Pan}}, \bibinfo {author} {\bibfnamefont {Y.}~\bibnamefont
  {Xu}}, \bibinfo {author} {\bibfnamefont {X.~G.}\ \bibnamefont {Luo}},
  \bibinfo {author} {\bibfnamefont {X.~H.}\ \bibnamefont {Chen}}, \ and\
  \bibinfo {author} {\bibfnamefont {S.~Y.}\ \bibnamefont {Li}},\ }\href
  {http://arxiv.org/abs/1403.0191} {\bibfield  {journal} {\bibinfo  {journal}
  {{arXiv}:1403.0191}\ } (\bibinfo {year} {2014})}\BibitemShut {NoStop}%
\bibitem [{\citenamefont {Shermadini}\ \emph {et~al.}(2012)\citenamefont
  {Shermadini}, \citenamefont {Luetkens}, \citenamefont {Maisuradze},
  \citenamefont {Khasanov}, \citenamefont {Bukowski}, \citenamefont {Klauss},\
  and\ \citenamefont {Amato}}]{shermadini_superfluid_2012}%
  \BibitemOpen
  \bibfield  {author} {\bibinfo {author} {\bibfnamefont {Z.}~\bibnamefont
  {Shermadini}}, \bibinfo {author} {\bibfnamefont {H.}~\bibnamefont
  {Luetkens}}, \bibinfo {author} {\bibfnamefont {A.}~\bibnamefont
  {Maisuradze}}, \bibinfo {author} {\bibfnamefont {R.}~\bibnamefont
  {Khasanov}}, \bibinfo {author} {\bibfnamefont {Z.}~\bibnamefont {Bukowski}},
  \bibinfo {author} {\bibfnamefont {H.-H.}\ \bibnamefont {Klauss}}, \ and\
  \bibinfo {author} {\bibfnamefont {A.}~\bibnamefont {Amato}},\ }\href
  {\doibase 10.1103/PhysRevB.86.174516} {\bibfield  {journal} {\bibinfo
  {journal} {Physical Review B}\ }\textbf {\bibinfo {volume} {86}},\ \bibinfo
  {pages} {174516} (\bibinfo {year} {2012})}\BibitemShut {NoStop}%
\bibitem [{\citenamefont {Sasmal}\ \emph {et~al.}(2008)\citenamefont {Sasmal},
  \citenamefont {Lv}, \citenamefont {Lorenz}, \citenamefont {Guloy},
  \citenamefont {Chen}, \citenamefont {Xue},\ and\ \citenamefont
  {Chu}}]{sasmal_superconducting_2008}%
  \BibitemOpen
  \bibfield  {author} {\bibinfo {author} {\bibfnamefont {K.}~\bibnamefont
  {Sasmal}}, \bibinfo {author} {\bibfnamefont {B.}~\bibnamefont {Lv}}, \bibinfo
  {author} {\bibfnamefont {B.}~\bibnamefont {Lorenz}}, \bibinfo {author}
  {\bibfnamefont {A.~M.}\ \bibnamefont {Guloy}}, \bibinfo {author}
  {\bibfnamefont {F.}~\bibnamefont {Chen}}, \bibinfo {author} {\bibfnamefont
  {Y.-Y.}\ \bibnamefont {Xue}}, \ and\ \bibinfo {author} {\bibfnamefont
  {C.-W.}\ \bibnamefont {Chu}},\ }\href {\doibase
  10.1103/PhysRevLett.101.107007} {\bibfield  {journal} {\bibinfo  {journal}
  {Physical Review Letters}\ }\textbf {\bibinfo {volume} {101}},\ \bibinfo
  {pages} {107007} (\bibinfo {year} {2008})}\BibitemShut {NoStop}%
\bibitem [{\citenamefont {Aftabuzzaman}\ and\ \citenamefont
  {Islam}(2010)}]{aftabuzzaman_new_2010}%
  \BibitemOpen
  \bibfield  {author} {\bibinfo {author} {\bibfnamefont {M.}~\bibnamefont
  {Aftabuzzaman}}\ and\ \bibinfo {author} {\bibfnamefont {A.~K. M.~A.}\
  \bibnamefont {Islam}},\ }\href {\doibase 10.1016/j.physc.2009.12.040}
  {\bibfield  {journal} {\bibinfo  {journal} {Physica C: Superconductivity}\
  }\textbf {\bibinfo {volume} {470}},\ \bibinfo {pages} {202} (\bibinfo {year}
  {2010})}\BibitemShut {NoStop}%
\bibitem [{\citenamefont {Hardy}\ \emph {et~al.}(2013)\citenamefont {Hardy},
  \citenamefont {B\"{o}hmer}, \citenamefont {Aoki}, \citenamefont {Burger},
  \citenamefont {Wolf}, \citenamefont {Schweiss}, \citenamefont {Heid},
  \citenamefont {Adelmann}, \citenamefont {Yao}, \citenamefont {Kotliar},
  \citenamefont {Schmalian},\ and\ \citenamefont
  {Meingast}}]{hardy_evidence_2013}%
  \BibitemOpen
  \bibfield  {author} {\bibinfo {author} {\bibfnamefont {F.}~\bibnamefont
  {Hardy}}, \bibinfo {author} {\bibfnamefont {A.~E.}\ \bibnamefont
  {B\"{o}hmer}}, \bibinfo {author} {\bibfnamefont {D.}~\bibnamefont {Aoki}},
  \bibinfo {author} {\bibfnamefont {P.}~\bibnamefont {Burger}}, \bibinfo
  {author} {\bibfnamefont {T.}~\bibnamefont {Wolf}}, \bibinfo {author}
  {\bibfnamefont {P.}~\bibnamefont {Schweiss}}, \bibinfo {author}
  {\bibfnamefont {R.}~\bibnamefont {Heid}}, \bibinfo {author} {\bibfnamefont
  {P.}~\bibnamefont {Adelmann}}, \bibinfo {author} {\bibfnamefont {Y.~X.}\
  \bibnamefont {Yao}}, \bibinfo {author} {\bibfnamefont {G.}~\bibnamefont
  {Kotliar}}, \bibinfo {author} {\bibfnamefont {J.}~\bibnamefont {Schmalian}},
  \ and\ \bibinfo {author} {\bibfnamefont {C.}~\bibnamefont {Meingast}},\
  }\href {\doibase 10.1103/PhysRevLett.111.027002} {\bibfield  {journal}
  {\bibinfo  {journal} {Physical Review Letters}\ }\textbf {\bibinfo {volume}
  {111}},\ \bibinfo {pages} {027002} (\bibinfo {year} {2013})}\BibitemShut
  {NoStop}%
\bibitem [{\citenamefont {Anderson}(1959)}]{anderson_theory_1959}%
  \BibitemOpen
  \bibfield  {author} {\bibinfo {author} {\bibfnamefont {P.~W.}\ \bibnamefont
  {Anderson}},\ }\href
  {http://www.sciencedirect.com/science/article/pii/0022369759900368}
  {\bibfield  {journal} {\bibinfo  {journal} {Journal of Physics and Chemistry
  of Solids}\ }\textbf {\bibinfo {volume} {11}},\ \bibinfo {pages} {26}
  (\bibinfo {year} {1959})}\BibitemShut {NoStop}%
\bibitem [{\citenamefont {Prozorov}\ \emph {et~al.}(2014)\citenamefont
  {Prozorov}, \citenamefont {Konczykowski}, \citenamefont {Tanatar},
  \citenamefont {Thaler}, \citenamefont {Bud'ko}, \citenamefont {Canfield},
  \citenamefont {Mishra},\ and\ \citenamefont
  {Hirschfeld}}]{prozorov_effect_2014}%
  \BibitemOpen
  \bibfield  {author} {\bibinfo {author} {\bibfnamefont {R.}~\bibnamefont
  {Prozorov}}, \bibinfo {author} {\bibfnamefont {M.}~\bibnamefont
  {Konczykowski}}, \bibinfo {author} {\bibfnamefont {M.~A.}\ \bibnamefont
  {Tanatar}}, \bibinfo {author} {\bibfnamefont {A.}~\bibnamefont {Thaler}},
  \bibinfo {author} {\bibfnamefont {S.~L.}\ \bibnamefont {Bud'ko}}, \bibinfo
  {author} {\bibfnamefont {P.~C.}\ \bibnamefont {Canfield}}, \bibinfo {author}
  {\bibfnamefont {V.}~\bibnamefont {Mishra}}, \ and\ \bibinfo {author}
  {\bibfnamefont {P.~J.}\ \bibnamefont {Hirschfeld}},\ }\href {\doibase
  10.1103/PhysRevX.4.041032} {\bibfield  {journal} {\bibinfo  {journal}
  {Physical Review X}\ }\textbf {\bibinfo {volume} {4}},\ \bibinfo {pages}
  {041032} (\bibinfo {year} {2014})}\BibitemShut {NoStop}%
\bibitem [{\citenamefont {Wang}\ \emph {et~al.}(2014)\citenamefont {Wang},
  \citenamefont {Zhou}, \citenamefont {Luo}, \citenamefont {Hong},
  \citenamefont {Yan}, \citenamefont {Ying}, \citenamefont {Cheng},
  \citenamefont {Ye}, \citenamefont {Xiang}, \citenamefont {Li},\ and\
  \citenamefont {Chen}}]{wang_anomalous_2014}%
  \BibitemOpen
  \bibfield  {author} {\bibinfo {author} {\bibfnamefont {A.~F.}\ \bibnamefont
  {Wang}}, \bibinfo {author} {\bibfnamefont {S.~Y.}\ \bibnamefont {Zhou}},
  \bibinfo {author} {\bibfnamefont {X.~G.}\ \bibnamefont {Luo}}, \bibinfo
  {author} {\bibfnamefont {X.~C.}\ \bibnamefont {Hong}}, \bibinfo {author}
  {\bibfnamefont {Y.~J.}\ \bibnamefont {Yan}}, \bibinfo {author} {\bibfnamefont
  {J.~J.}\ \bibnamefont {Ying}}, \bibinfo {author} {\bibfnamefont
  {P.}~\bibnamefont {Cheng}}, \bibinfo {author} {\bibfnamefont {G.~J.}\
  \bibnamefont {Ye}}, \bibinfo {author} {\bibfnamefont {Z.~J.}\ \bibnamefont
  {Xiang}}, \bibinfo {author} {\bibfnamefont {S.~Y.}\ \bibnamefont {Li}}, \
  and\ \bibinfo {author} {\bibfnamefont {X.~H.}\ \bibnamefont {Chen}},\ }\href
  {\doibase 10.1103/PhysRevB.89.064510} {\bibfield  {journal} {\bibinfo
  {journal} {Physical Review B}\ }\textbf {\bibinfo {volume} {89}},\ \bibinfo
  {pages} {064510} (\bibinfo {year} {2014})}\BibitemShut {NoStop}%
\bibitem [{\citenamefont {Kirshenbaum}\ \emph {et~al.}(2012)\citenamefont
  {Kirshenbaum}, \citenamefont {Saha}, \citenamefont {Ziemak}, \citenamefont
  {Drye},\ and\ \citenamefont {Paglione}}]{kirshenbaum_universal_2012}%
  \BibitemOpen
  \bibfield  {author} {\bibinfo {author} {\bibfnamefont {K.}~\bibnamefont
  {Kirshenbaum}}, \bibinfo {author} {\bibfnamefont {S.~R.}\ \bibnamefont
  {Saha}}, \bibinfo {author} {\bibfnamefont {S.}~\bibnamefont {Ziemak}},
  \bibinfo {author} {\bibfnamefont {T.}~\bibnamefont {Drye}}, \ and\ \bibinfo
  {author} {\bibfnamefont {J.}~\bibnamefont {Paglione}},\ }\href@noop {}
  {\bibfield  {journal} {\bibinfo  {journal} {Physical Review B}\ }\textbf
  {\bibinfo {volume} {86}},\ \bibinfo {pages} {140505} (\bibinfo {year}
  {2012})}\BibitemShut {NoStop}%
\bibitem [{\citenamefont {Maiti}\ \emph {et~al.}(2012)\citenamefont {Maiti},
  \citenamefont {Korshunov},\ and\ \citenamefont {Chubukov}}]{maiti_gap_2012}%
  \BibitemOpen
  \bibfield  {author} {\bibinfo {author} {\bibfnamefont {S.}~\bibnamefont
  {Maiti}}, \bibinfo {author} {\bibfnamefont {M.~M.}\ \bibnamefont
  {Korshunov}}, \ and\ \bibinfo {author} {\bibfnamefont {A.~V.}\ \bibnamefont
  {Chubukov}},\ }\href@noop {} {\bibfield  {journal} {\bibinfo  {journal}
  {Physical Review B}\ }\textbf {\bibinfo {volume} {85}},\ \bibinfo {pages}
  {014511} (\bibinfo {year} {2012})}\BibitemShut {NoStop}%
\bibitem [{\citenamefont {Rotter}\ \emph {et~al.}(2008)\citenamefont {Rotter},
  \citenamefont {Pangerl}, \citenamefont {Tegel},\ and\ \citenamefont
  {Johrendt}}]{rotter_superconductivity_2008}%
  \BibitemOpen
  \bibfield  {author} {\bibinfo {author} {\bibfnamefont {M.}~\bibnamefont
  {Rotter}}, \bibinfo {author} {\bibfnamefont {M.}~\bibnamefont {Pangerl}},
  \bibinfo {author} {\bibfnamefont {M.}~\bibnamefont {Tegel}}, \ and\ \bibinfo
  {author} {\bibfnamefont {D.}~\bibnamefont {Johrendt}},\ }\href {\doibase
  10.1002/anie.200803641} {\bibfield  {journal} {\bibinfo  {journal}
  {Angewandte Chemie International Edition}\ }\textbf {\bibinfo {volume}
  {47}},\ \bibinfo {pages} {7949} (\bibinfo {year} {2008})}\BibitemShut
  {NoStop}%
\bibitem [{\citenamefont {Miyoshi}\ \emph {et~al.}(2013)\citenamefont
  {Miyoshi}, \citenamefont {Kojima}, \citenamefont {Ogawa}, \citenamefont
  {Shimojo},\ and\ \citenamefont {Takeuchi}}]{miyoshi_superconductivity_2013}%
  \BibitemOpen
  \bibfield  {author} {\bibinfo {author} {\bibfnamefont {K.}~\bibnamefont
  {Miyoshi}}, \bibinfo {author} {\bibfnamefont {E.}~\bibnamefont {Kojima}},
  \bibinfo {author} {\bibfnamefont {S.}~\bibnamefont {Ogawa}}, \bibinfo
  {author} {\bibfnamefont {Y.}~\bibnamefont {Shimojo}}, \ and\ \bibinfo
  {author} {\bibfnamefont {J.}~\bibnamefont {Takeuchi}},\ }\href {\doibase
  10.1103/PhysRevB.87.235111} {\bibfield  {journal} {\bibinfo  {journal}
  {Physical Review B}\ }\textbf {\bibinfo {volume} {87}},\ \bibinfo {pages}
  {235111} (\bibinfo {year} {2013})}\BibitemShut {NoStop}%
\bibitem [{\citenamefont {Kretzschmar}\ \emph {et~al.}(2013)\citenamefont
  {Kretzschmar}, \citenamefont {Muschler}, \citenamefont {B\"{o}hm},
  \citenamefont {Baum}, \citenamefont {Hackl}, \citenamefont {Wen},
  \citenamefont {Tsurkan}, \citenamefont {Deisenhofer},\ and\ \citenamefont
  {Loidl}}]{kretzschmar_raman-scattering_2013}%
  \BibitemOpen
  \bibfield  {author} {\bibinfo {author} {\bibfnamefont {F.}~\bibnamefont
  {Kretzschmar}}, \bibinfo {author} {\bibfnamefont {B.}~\bibnamefont
  {Muschler}}, \bibinfo {author} {\bibfnamefont {T.}~\bibnamefont {B\"{o}hm}},
  \bibinfo {author} {\bibfnamefont {A.}~\bibnamefont {Baum}}, \bibinfo {author}
  {\bibfnamefont {R.}~\bibnamefont {Hackl}}, \bibinfo {author} {\bibfnamefont
  {H.-H.}\ \bibnamefont {Wen}}, \bibinfo {author} {\bibfnamefont
  {V.}~\bibnamefont {Tsurkan}}, \bibinfo {author} {\bibfnamefont
  {J.}~\bibnamefont {Deisenhofer}}, \ and\ \bibinfo {author} {\bibfnamefont
  {A.}~\bibnamefont {Loidl}},\ }\href {\doibase 10.1103/PhysRevLett.110.187002}
  {\bibfield  {journal} {\bibinfo  {journal} {Physical Review Letters}\
  }\textbf {\bibinfo {volume} {110}},\ \bibinfo {pages} {187002} (\bibinfo
  {year} {2013})}\BibitemShut {NoStop}%
\bibitem [{\citenamefont {Liu}\ \emph {et~al.}(2010)\citenamefont {Liu},
  \citenamefont {Kondo}, \citenamefont {Fernandes}, \citenamefont {Palczewski},
  \citenamefont {Mun}, \citenamefont {Ni}, \citenamefont {Thaler},
  \citenamefont {Bostwick}, \citenamefont {Rotenberg}, \citenamefont
  {Schmalian}, \citenamefont {Bud'ko}, \citenamefont {Canfield},\ and\
  \citenamefont {Kaminski}}]{liu_evidence_2010}%
  \BibitemOpen
  \bibfield  {author} {\bibinfo {author} {\bibfnamefont {C.}~\bibnamefont
  {Liu}}, \bibinfo {author} {\bibfnamefont {T.}~\bibnamefont {Kondo}}, \bibinfo
  {author} {\bibfnamefont {R.~M.}\ \bibnamefont {Fernandes}}, \bibinfo {author}
  {\bibfnamefont {A.~D.}\ \bibnamefont {Palczewski}}, \bibinfo {author}
  {\bibfnamefont {E.~D.}\ \bibnamefont {Mun}}, \bibinfo {author} {\bibfnamefont
  {N.}~\bibnamefont {Ni}}, \bibinfo {author} {\bibfnamefont {A.~N.}\
  \bibnamefont {Thaler}}, \bibinfo {author} {\bibfnamefont {A.}~\bibnamefont
  {Bostwick}}, \bibinfo {author} {\bibfnamefont {E.}~\bibnamefont {Rotenberg}},
  \bibinfo {author} {\bibfnamefont {J.}~\bibnamefont {Schmalian}}, \bibinfo
  {author} {\bibfnamefont {S.~L.}\ \bibnamefont {Bud'ko}}, \bibinfo {author}
  {\bibfnamefont {P.~C.}\ \bibnamefont {Canfield}}, \ and\ \bibinfo {author}
  {\bibfnamefont {A.}~\bibnamefont {Kaminski}},\ }\href {\doibase
  10.1038/nphys1656} {\bibfield  {journal} {\bibinfo  {journal} {Nature
  Physics}\ }\textbf {\bibinfo {volume} {6}},\ \bibinfo {pages} {419} (\bibinfo
  {year} {2010})}\BibitemShut {NoStop}%
\bibitem [{\citenamefont {Kogan}\ and\ \citenamefont
  {Prozorov}(2012)}]{kogan_orbital_2012}%
  \BibitemOpen
  \bibfield  {author} {\bibinfo {author} {\bibfnamefont {V.~G.}\ \bibnamefont
  {Kogan}}\ and\ \bibinfo {author} {\bibfnamefont {R.}~\bibnamefont
  {Prozorov}},\ }\href {\doibase 10.1088/0034-4885/75/11/114502} {\bibfield
  {journal} {\bibinfo  {journal} {Reports on Progress in Physics}\ }\textbf
  {\bibinfo {volume} {75}},\ \bibinfo {pages} {114502} (\bibinfo {year}
  {2012})}\BibitemShut {NoStop}%
\bibitem [{\citenamefont {Thomale}\ \emph {et~al.}(2011)\citenamefont
  {Thomale}, \citenamefont {Platt}, \citenamefont {Hanke}, \citenamefont {Hu},\
  and\ \citenamefont {Bernevig}}]{thomale_exotic_2011}%
  \BibitemOpen
  \bibfield  {author} {\bibinfo {author} {\bibfnamefont {R.}~\bibnamefont
  {Thomale}}, \bibinfo {author} {\bibfnamefont {C.}~\bibnamefont {Platt}},
  \bibinfo {author} {\bibfnamefont {W.}~\bibnamefont {Hanke}}, \bibinfo
  {author} {\bibfnamefont {J.}~\bibnamefont {Hu}}, \ and\ \bibinfo {author}
  {\bibfnamefont {B.~A.}\ \bibnamefont {Bernevig}},\ }\href {\doibase
  10.1103/PhysRevLett.107.117001} {\bibfield  {journal} {\bibinfo  {journal}
  {Physical Review Letters}\ }\textbf {\bibinfo {volume} {107}},\ \bibinfo
  {pages} {117001} (\bibinfo {year} {2011})}\BibitemShut {NoStop}%
\end{thebibliography}%

\end{document}